\documentstyle[12pt,epsf]{article}
\def\fo{\hbox{{1}\kern-.25em\hbox{l}}}
\def\fnote#1#2{\begingroup\def\thefootnote{#1}\footnote{#2}\addtocounter
{footnote}{-1}\endgroup}

\renewcommand{\thefootnote}{\fnsymbol{footnote}}

\def\beq{\begin{equation}}
\def\eeq{\end{equation}}
\def\eq{\end{equation}}
\def\to{\rightarrow}
\def\mxc#1{m_{\tilde \chi^\pm_{#1}}}
\def\msq{m_{\tilde q}}
\def\mst#1{m_{\tilde t_{#1}}}
\def\fyg#1{f^{(#1)}_{\gamma ,g}}
\def\fy#1{f^{(#1)}_\gamma}
\def\fg#1{f^{(#1)}_g}

\def\bsg{\ifmmode B\to X_s\gamma\else $B\to X_s\gamma$\fi}
\def\bsll{\ifmmode B\to X_s\ell^+\ell^-\else $B\to X_s\ell^+\ell^-$\fi}
\def\bstt{\ifmmode B\to X_s\tau^+\tau^-\else $B\to X_s\tau^+\tau^-$\fi}
\def\shat{\ifmmode \hat{s}\else $\hat{s}$\fi}

\newcommand{\newc}{\newcommand}

\newc{\lcal}{\int {\cal L}dt}

\newc{\mHpm}{m_{H^\pm}}
\newc{\gsim}{\lower.7ex\hbox{$\;\stackrel{\textstyle>}{\sim}\;$}}
\newc{\lsim}{\lower.7ex\hbox{$\;\stackrel{\textstyle<}{\sim}\;$}}
\newc{\ie}{{\it i.e.}}          
\newc{\etal}{{\it et al.}}
\newc{\eg}{{\it e.g.}}          
\newc{\kev}{\hbox{\rm\,keV}}            
\newc{\mev}{\hbox{\rm\,MeV}}            
\newc{\gev}{\hbox{\rm\,GeV}}            
\newc{\tev}{\hbox{\rm\,TeV}}
\newc{\xpb}{\hbox{\rm\, pb}}
\newc{\xfb}{\hbox{\rm\, fb}}
\def\order#1{{\cal O}(#1)}
\newc{\mtop}{m_t}
\newc{\mbot}{m_b}
\newc{\mz}{m_Z}
\newc{\mw}{M_W}
\newc{\alphasmz}{\alpha_s(m_Z^2)}
\newc{\swsq}{\sin^2\theta_W}
\newc{\tw}{\tan\theta_W}
\newc{\cw}{\cos\theta_W}
\newc{\sw}{\sin\theta_W}
\newc{\BR}{\hbox{\rm BR}}
\newc{\zbb}{Z\to b\bar}
\newc{\Gb}{\Gamma (Z\to b\bar b)}
\newc{\Gh}{\Gamma (Z\to \hbox{\rm hadrons})}
\newc{\rbsm}{R_b^\hbox{\rm sm}}
\newc{\rbsusy}{R_b^\hbox{\rm susy}}
\newc{\drb}{\delta R_b}

\newc{\sgn}{\mbox{sgn}}
\newc{\tbeta}{\tan\beta}
\newc{\uL}{{\tilde u_L}}
\newc{\uR}{{\tilde u_R}}
\newc{\cL}{{\tilde c_L}}
\newc{\cR}{{\tilde c_R}}
\newc{\tL}{{\tilde t_L}}
\newc{\tR}{{\tilde t_R}}
\newc{\dL}{{\tilde d_L}}
\newc{\dR}{{\tilde d_R}}
\newc{\sL}{{\tilde s_L}}
\newc{\sR}{{\tilde s_R}}
\newc{\bL}{{\tilde b_L}}
\newc{\bR}{{\tilde b_R}}
\newc{\eL}{{\tilde e_L}}
\newc{\eR}{{\tilde e_R}}
\newc{\mhp}{m_{H^\pm}}
\newc{\mhalf}{m_{1/2}}

\newc{\lR}{\tilde{l}_R}
\newc{\lL}{\tilde{l}_L}
\newc{\nL}{\tilde{\nu}_L}
\newc{\na}{\chi^0_1}
\newc{\nb}{\chi^0_2}
\newc{\nc}{\chi^0_3}
\newc{\nd}{\chi^0_4}
\newc{\ca}{\chi^{\pm}_1}
\newc{\cb}{\chi^{\pm}_2}
\newc{\camp}{\chi^\mp_1}
\newc{\cbmp}{\chi^\mp_1}
\newc{\capos}{\chi^{+}_1}
\newc{\caneg}{\chi^{-}_1}

\def\NPB#1#2#3{Nucl. Phys. B {\bf #1} (19#2) #3}
\def\PLB#1#2#3{Phys. Lett. B {\bf #1} (19#2) #3}

\def\PRD#1#2#3{Phys. Rev. D {\bf #1} (19#2) #3}
\def\PRL#1#2#3{Phys. Rev. Lett. {\bf#1} (19#2) #3}

\def\ZPC#1#2#3{Zeit. f\"ur Physik C {\bf #1} (19#2) #3}

\def\beq{\begin{equation}}
\def\eeq{\end{equation}}
\def\bea{\begin{eqnarray*}}
\def\eea{\end{eqnarray*}}
\def\slashchar#1{\setbox0=\hbox{$#1$}           
   \dimen0=\wd0                                 
   \setbox1=\hbox{/} \dimen1=\wd1               
   \ifdim\dimen0>\dimen1                        
      \rlap{\hbox to \dimen0{\hfil/\hfil}}      
      #1                                        
   \else                                        
      \rlap{\hbox to \dimen1{\hfil$#1$\hfil}}   
      /                                         
   \fi}                                         %
\catcode`@=11
\long\def\@caption#1[#2]#3{\par\addcontentsline{\csname
  ext@#1\endcsname}{#1}{\protect\numberline{\csname
  the#1\endcsname}{\ignorespaces #2}}\begingroup
    \small
    \@parboxrestore
    \@makecaption{\csname fnum@#1\endcsname}{\ignorespaces #3}\par
  \endgroup}
\catcode`@=12

\def\jfig#1#2#3{
 \begin{figure}
 \centering
 \epsfysize=2.4in
 \hspace*{0in}
 \epsffile{#2}
 \caption{#3}
 \label{#1}
 \end{figure}}

\begin{document}

\begin{titlepage}

\begin{flushright}
SLAC-PUB-7290 \\
hep-ph/9610323\\
October 1996
\end{flushright}


\huge
\begin{center}
Searching for supersymmetry \\
 in rare $B$ decays
\end{center}

\large

\vspace{.15in}
\begin{center}

JoAnne L. Hewett, James D. Wells \\

\vspace{.1in}
{\it Stanford Linear Accelerator Center \\
Stanford University, Stanford, CA 94309\fnote{\dagger}{Work 
supported by the Department of Energy
under contract DE-AC03-76SF00515.}\\}

\end{center}
 
 
\vspace{0.15in}

\normalsize

\begin{abstract}

We quantify the ability of $B$-Factories to observe supersymmetric
contributions to the rare decays $B \to X_s \gamma$ and 
$B \to X_s l^+l^-$.  A global
fit to the Wilson coefficients which contribute to these decays is performed
from Monte Carlo generated data on $B(B \to X_s \gamma)$ and
the kinematic distributions associated with the final state lepton
pair in $B\to X_sl^+l^-$.
This fit is then compared to
supersymmetric predictions.  Evaluation of the Wilson coefficients
is carried out with several different patterns of the superpartner spectrum.
We find that $B$-Factories will be able to probe
regions of the SUSY parameter space not accessible to LEP~II, the Tevatron,
and perhaps the LHC.
We also employ the recent NLO calculation of the
matrix elements for \bsg\ and find the bound $m_{H^\pm}>300\gev$ in
two-Higgs-doublet models using present data.

\end{abstract}

\end{titlepage}

\baselineskip=18pt



\section{Introduction}

Softly broken supersymmetry (SUSY) is a 
decoupling theory, thus making it a challenge to search for its effects
through indirect methods.  
When competing with standard model tree-level
processes, the relative shift in an electroweak observable with respect to the 
standard model (SM) value should not be expected to exceed much more than
$(\alpha_2/2\pi)\mw^2/\tilde m^2$.
In the post-LEP~II and Tevatron era, 
if supersymmetry has not been directly observed,
then $\tilde m\gsim \mw$, and so the relative shift expected
in standard model observables from virtual supersymmetry is $\lsim 0.5\%$.  
Although the $1\sigma$ bounds on
$\swsq$ are approaching this level from the analysis of SLC/LEP data,
a more statistically significant result would be difficult to obtain
given the current data sets available.  

Another approach to indirect searches of supersymmetry is to measure
observables where supersymmetry and the standard model arise at the
same order in perturbation theory. In this case, the SUSY
contributions do not suffer an extra $\alpha /4\pi$ reduction compared
to the standard model amplitudes.  The relative ratio between the 
lowest order standard model amplitudes and supersymmetric partner amplitudes
could then be ${\cal O}(1)$ if $\tilde m\simeq \mw$.
Rare $B$-decay measurements provide an opportunity for discovering
indirect effects of supersymmetry by this second approach.  

Two problems in the past have marred the attempts to use rare $B$ decays
as a good probe for physics beyond the standard model. The first is limited
statistics, or rather the 
number of $B$ mesons available in data sets which can be
used to study
and obtain good precision on low branching fraction modes.  The 
$B$ factories presently under construction, which will collect some 
$10^{7-8}$ $B$ mesons per year, will
alleviate this issue.  The second difficulty is theoretical.  Since all
the processes occur near $5\gev$ the uncertainties in the 
strong interactions can hide even 
$\order{1}$ effects in the electroweak contributions.  However, this problem
diminishes 
significantly~\cite{buraswarsaw,buras94:374,greub:96}
with a complete program of 
next-to-leading order (NLO)
computations of the QCD corrections to rare $B$ decays.
For the processes we will consider here, \bsg\ and \bsll, these higher order
calculations have essentially been 
completed recently~\cite{buras:95,greub:96,misiak:96}.
The inclusive decay \bsg\ has been observed by CLEO~\cite{cleo:94} 
with a branching 
fraction of $(2.32\pm 0.57\pm 0.35)\times 10^{-4}$ and $95\%$ C.L.
bounds of $1\times 10^{-4}< B(\bsg)<4.2\times 10^{-4}$.  
Meanwhile, experiments at $e^+e^-$
and hadron colliders are closing in on the observation\cite{bsll} of the 
exclusive modes $B\to K^{(*)}\ell^+\ell^-$ with $\ell=e$ and  $\mu$, 
respectively.  Once this decay is observed, the utilization of the
kinematic distributions of the $\ell^+\ell^-$ pair, such as
the lepton pair invariant mass distribution and forward
backward asymmetry~\cite{ali,ali2}, and the tau polarization 
asymmetry~\cite{hewett} 
in $B\to X_s\tau^+\tau^-$, together with $B(\bsg)$ will provide a stringent
test of the SM.

The outline of this paper is as follows.
In section 2 we calculate the ability of future $B$-factories to experimentally
determine the magnitude and sign of the relevant
Wilson coefficients in the rare $B$ decay interaction Hamiltonian 
using a global fit procedure.  We find
that the sensitivity for new physics will be substantially increased
beyond what is currently possible.  In section 3 we apply these results
to supersymmetry and estimate the sensitivity to high supersymmetric
mass scales.  We also reexamine the constraints on the $H^\pm$ sector from
\bsg, in light of the recent NLO computations.
And in the final section we discuss our conclusions.

\section{Determination of the Wilson coefficients}

The effective field theory for the decays \bsg\ and \bsll, which incorporates 
the QCD corrections, is governed by the Hamiltonian
\begin{equation}
{\cal H}_{eff}={-4G_F\over\sqrt 2}V_{tb}V^*_{ts}\sum_{i=1}^{10}C_i(\mu)
{\cal O_i}(\mu)\,,
\label{effham}
\end{equation}
where the ${\cal O}_i$ are a complete set of renormalized operators of
dimension six or less which mediate $b\to s$ transitions.  These operators
are catalogued in, \eg , Ref.~\cite{grinstein:89}.  The $C_i$ represent
the corresponding Wilson coefficients which are evaluated perturbatively
at the electroweak scale where the matching conditions are imposed and
then evolved down to the renormalization scale $\mu\approx  m_b$.  We note that
the magnetic and chromomagnetic dipole operators, ${\cal O}_{7,8}$, contain 
explicit mass factors which must also be renormalized.

For \bsll\ this formalism leads to the physical decay amplitude (neglecting 
the strange quark mass)
\begin{eqnarray}
{\cal M}(\bsll) & = & {\sqrt 2G_F\alpha\over \pi}V_{tb}V^*_{ts}
\left[ C_9^{eff}\bar s_L\gamma_\mu b_L\bar\ell\gamma^\mu\ell +
C_{10}\bar s_L\gamma_\mu b_L\bar\ell\gamma^\mu\gamma_5\ell \right. \nonumber\\
& & \quad\quad \left.
-2C^{eff}_7m_b\bar s_Li\sigma_{\mu\nu}{q^\nu\over q^2}b_R\bar\ell\gamma^\mu\ell
\right] \,, 
\end{eqnarray}
where $q^2$ represents the momentum transferred to the lepton pair.  The
expressions for $C_i(M_W)$ are given by the Inami-Lim functions\cite{inami}.
A NLO analysis
for this decay has recently been performed\cite{buras:95}, where it is
stressed that a scheme independent result can only be obtained by including
the leading and next-to-leading logarithmic corrections to
$C_9(\mu)$ while retaining
only the leading logarithms in the remaining Wilson coefficients.  The
residual leading $\mu$ dependence in $C_9(\mu)$ is cancelled by that
contained in the matrix element of ${\cal O}_9$.  
The combination yields an effective value of $C_9$ given by,
\begin{equation}
C^{eff}_9 (\hat s)= C_9(\mu)\eta(\shat)+Y(\shat) \,,
\end{equation}
with $Y(\shat)$ being the one-loop matrix element of ${\cal O}_9$,
$\eta(\shat)$ represents the single gluon corrections to this matrix element, 
and $\shat\equiv q^2/m_b^2$ is the scaled momentum transferred to the lepton
pair.  The effective value for $C^{eff}_7(\mu )$ refers to the leading
order scheme independent result obtained by Buras \etal~\cite{buras94:374}.
The corresponding formulae for $C_i(\mu)$,
$Y(\shat)$ and $\eta(\shat)$ are collected in 
Refs. \cite{buras:95,grinstein:89}.
The operator ${\cal O}_{10}$ does not renormalize and hence
its corresponding coefficient does not depend on 
the value of $\mu$ (except for the
$\mu$ dependence associated with the definition of the top-quark mass).
The numerical estimates (in the naive dimensional regularization (NDR) 
scheme) for these
coefficients are then (taking $m^{pole}_b=4.87\gev$, $m_t^{phys}=175\gev$, and 
$\alpha_s (M_Z)=0.118$)
\begin{eqnarray}
C^{eff}_7(\mu=m_b~^{-m_b/2}_{+m_b}) & = & 
         -0.312_{+0.034}^{-0.059} \,,\nonumber \\
C_9(\mu=m_b~^{-m_b/2}_{+m_b}) & = & 4.21^{+0.31}_{-0.40} \,, 
\end{eqnarray}
and
\begin{equation}
C_{10}(\mu) = -4.55 \,.
\end{equation}
The reduced scale dependence of the NLO versus the LO corrected coefficients
is reflected in the deviations $\Delta C_9(\mu)\lsim\pm 10\%$ and 
$\Delta C^{eff}_7(\mu)\approx\pm 20\%$ as $\mu$ is varied in the range
$m_b/2\leq\mu\leq 2m_b$.  We find that the coefficients are much less
sensitive to the values of the remaining input parameters, with $\Delta
C_9(m_b),\Delta C^{eff}_7(m_b)\lsim 3\%$, varying 
$\alpha_s(M_Z)=0.118\pm 0.003$
\cite{pdg,schmelling}, 
and $m_t^{phys}=175\pm 6\gev$ \cite{cdfd0} corresponding to
$m_t(m_t)=166\pm 6\gev$.
The resulting inclusive branching fractions (which are computed by scaling
the width for \bsll\ to that for $B$ semi-leptonic decay)  are found
to be $(6.25^{+1.04}_{-0.93})\times 10^{-6}$,
$(5.73^{+0.75}_{-0.78})\times 10^{-6}$, 
and $(3.24^{+0.44}_{-0.54})\times 10^{-7}$
for $\ell=e, \mu$, and $\tau$, respectively, taking into account the
above input parameter ranges, as well as 
$B_{sl}\equiv B(B\to X\ell\nu)=(10.23\pm 0.39)\%$~\cite{richman}, and
$m_c/m_b=0.29\pm 0.02$ \cite{greub:96,pdg}.  
There are also long distance resonance contributions 
to \bsll, arising from $B\to K^{(*)}\psi^{(')}\to K^{(*)}\ell^+\ell^-$.
These appear as an effective $(\bar s_L\gamma_\mu b_L)(\bar\ell\gamma_\mu\ell)$
interaction and are incorporated into $C_9^{eff}$ via the modification
$Y(\shat)\to Y'(\shat)\equiv Y(\shat)+Y_{res}(\shat)$, where $Y_{res}(\shat)$
is given in Ref. \cite{desh:89}.  These pole contributions lead to a 
significant
interference between the dispersive part of the resonance and the short
distance contributions.  However, suitable cuts on the lepton pair mass
spectrum can cleanly separate the short distance physics from the resonance
contributions.

The operator basis for the decay \bsg\ contains the first eight operators
in the effective Hamiltonian of Eq.~(\ref{effham}).  The leading logarithmic 
QCD corrections to the decay width have been completely resummed, 
but lead to a sizeable $\mu$ dependence of the branching fraction 
(as demonstrated above with the large value of $\Delta C^{eff}_7$),
and hence it is essential to include the next-to-leading order corrections to
reduce the theoretical uncertainty.  In this case, the calculation 
of the perturbative QCD corrections involves several steps, requiring 
corrections to both the Wilson coefficients and the matrix elements of the 
operators in Eq.~(\ref{effham}) in order to ensure a scheme independent result.
For the matrix elements, this includes the QCD bremsstrahlung 
corrections\cite{greub:91} $b\to s\gamma+g$, and the NLO virtual corrections
which have recently been completed in both the NDR
and 't Hooft-Veltman schemes~\cite{greub:96}.  
Summing these contributions to the matrix elements
and expanding them around $\mu=m_b$, one arrives at the decay 
amplitude
\begin{equation}
{\cal M}(b\to s\gamma) = -{4G_FV_{tb}V^*_{ts}\over\sqrt 2}D\langle
s\gamma|{\cal O}_7(m_b)|b\rangle_{tree} \,,
\end{equation}
with
\begin{equation}
\label{dc7eq}
D=C_7^{eff}(\mu)+{\alpha_s(m_b)\over 4\pi}\left( C_i^{(0)eff}(\mu)
\gamma_{i7}^{(0)eff}\log {m_b\over\mu} + C_i^{(0)eff}r_i\right) \,.
\end{equation}
Here, the quantities $\gamma_{i7}^{(0)eff}$ are the entries of the effective
leading order anomalous dimension matrix, and the $r_i$ are computed in Greub
\etal~\cite{greub:96}, for $i=2,7,8$.  The first term in Eq.~\ref{dc7eq}, 
$C_7^{eff}(\mu)$, must be computed
at NLO precision, while it is consistent to use the leading order values
of the other coefficients.  The explicit logarithms $\alpha_s(m_b)\log 
({m_b/\mu})$ in the equation
are cancelled by the $\mu$ dependence of $C_7^{(0)eff}(\mu)$.
This feature significantly reduces the scale dependence of the resulting
branching fraction.  The contribution to the inclusive width including these
virtual corrections is then
\begin{equation}
\Gamma^{virt}_{NLO}(B\to X_s\gamma)
 ={m^5_{b,pole}G_F^2\alpha_{em}|V_{tb}V^*_{ts}|^2\over 32\pi^4} F|D|^2 \,,
\end{equation}
where the factor $F=m_b^2(m_b)/m^2_{b,pole}=1-8\alpha_s(m_b)/3\pi$ arises
from the mass factor present in the magnetic dipole operator. This should
be compared to the familiar leading order result (which omits
the virtual corrections to $\langle {\cal O}_7\rangle$)
\beq
\Gamma (B\to X_s\gamma) = \frac{m_{b,pole}^5G^2_F\alpha_{em}}{32\pi^4}
 |V_{tb}V^*_{ts}|^2 |C_7^{eff}(\mu )|^2.
\eeq

For the Wilson coefficients, the NLO result entails
the computation of the ${\cal O}(\alpha_s)$ terms in the matching conditions,
and the renormalization group evolution of the $C_i(\mu)$ must be computed
using the ${\cal O}(\alpha_s^2)$ anomalous dimension matrix.  The former step
has been completed\cite{yao}, but the latter step is quite difficult since
some entries in the matrix have to be extracted from three-loop diagrams.
Nonetheless, preliminary NLO results for these anomalous dimensions have 
recently been reported~\cite{misiak:96}, with
the conclusion being that in the NDR scheme the NLO
correction to $C^{eff}_7(\mu)$ 
is small.  Therefore, a good approximation for the
inclusive width is obtained by employing the leading order expression for 
$C^{eff}_7(\mu )$, 
with the understanding that this introduces a small inherent 
uncertainty in the calculation.

The total inclusive width is then given by the sum of the virtual and
bremsstrahlung corrections, $\Gamma(\bsg)=\Gamma^{virt}+\Gamma^{brems}$, where
$\Gamma^{brems}$ is given in Greub \etal ~\cite{greub:91,greub:96}, and the
branching fraction is calculated by scaling to the semi-leptonic decay rate.
The leading order power corrections in the heavy quark expansion are identical
for \bsg\ and $B\to X\ell\nu$, and hence cancel in the ratio~\cite{hqet}.  
This 
allows us to approximate $\Gamma(\bsg)$ with the perturbatively calculable
free quark decay rate.  For $m^{phys}_t=175\pm 6\gev$, $m_b/2\leq\mu\leq 2m_b$,
$\alpha_s=0.118\pm 0.003$, $B_{sl}=(10.23\pm 0.39)\%$, and
$m_c/m_b=0.29 \pm 0.02$, we find the branching fraction
\begin{equation}
B(\bsg)=(3.25\pm 0.30 \pm 0.40)\times 10^{-4} \,,
\end{equation}
where the first error corresponds to the combined uncertainty associated with 
the value of $m_t$ and $\mu$, and the second error represents the uncertainty
from the other parameters.  This is well within the range observed by CLEO.
In Fig.~\ref{fig1} we display our results for
$B(B\to X_s\gamma )$ as a function of the top mass.
The dashed lines indicate the error in the branching ratio if
we fix $\mu =m_b$ and vary all the other parameters over their allowed ranges
given above. The solid lines indicate the error for
$m_b/2< \mu < 2m_b$ with all other parameters fixed to their central values.
This visually demonstrates that the error in the theoretical calculation
of $B\to X_s\gamma$ is not overwhelmed by the scale uncertainty; other 
uncertainties are now comparable.

\jfig{fig1}
{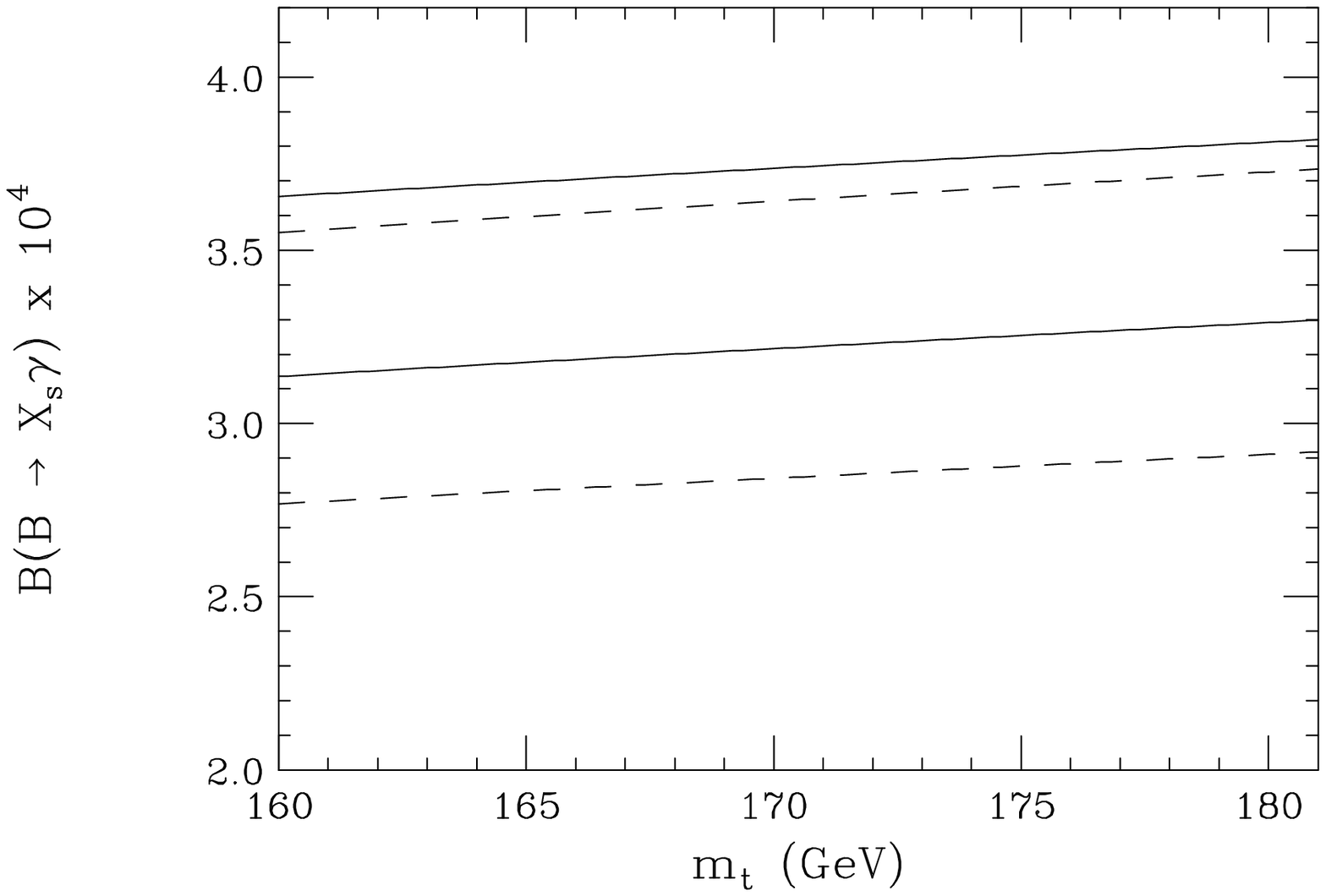}{The branching ratio of $B\to X_s\gamma$ versus $m_{t}$.
The dashed lines indicate the error in the branching ratio if
we fix $\mu =m_b$ and vary all the other parameters over their allowed ranges:
$\alpha_s(M_Z)=0.118\pm 0.003$, $B_{sl}=10.23\pm 0.39\%$, 
and $m_c/m_b=0.29\pm 0.02$.  The solid lines indicate the error for
$m_b/2< \mu < 2m_b$ and all other parameters fixed to their central values.}

Measurements of $B(\bsg)$ alone constrain the magnitude, but not the sign,
of $C^{eff}_7(\mu)$.  We can write the coefficients
at the matching scale in the form $C_i(M_W)=C_i^{SM}(M_W)+C_i^{new}(M_W)$,
where $C_i^{new}(M_W)$ clearly represents the contributions from new
interactions.  Due to operator mixing, \bsg\ then limits the 
possible values for $C_i^{new}(M_W)$ for $i=7,8$.  These bounds are
summarized in Fig.~\ref{fig2}.
Here, the solid bands correspond to the
constraints obtained from the current CLEO measurement,
taking into account the variation of the renormalization scale
$m_b/2 \leq \mu \leq 2m_b$, as well as the allowed ranges of
the other input parameters.  The dashed bands represent the constraints
when the scale is fixed to $\mu =m_b$.  We note that large values of
$C_8^{new}(\mw)$ are allowed even in the region where
$C_7^{new}(\mw)\simeq 0$.  Experimental bounds on the decay 
$b\to sg$ are needed to constraint $C_8$.

\jfig{fig2}{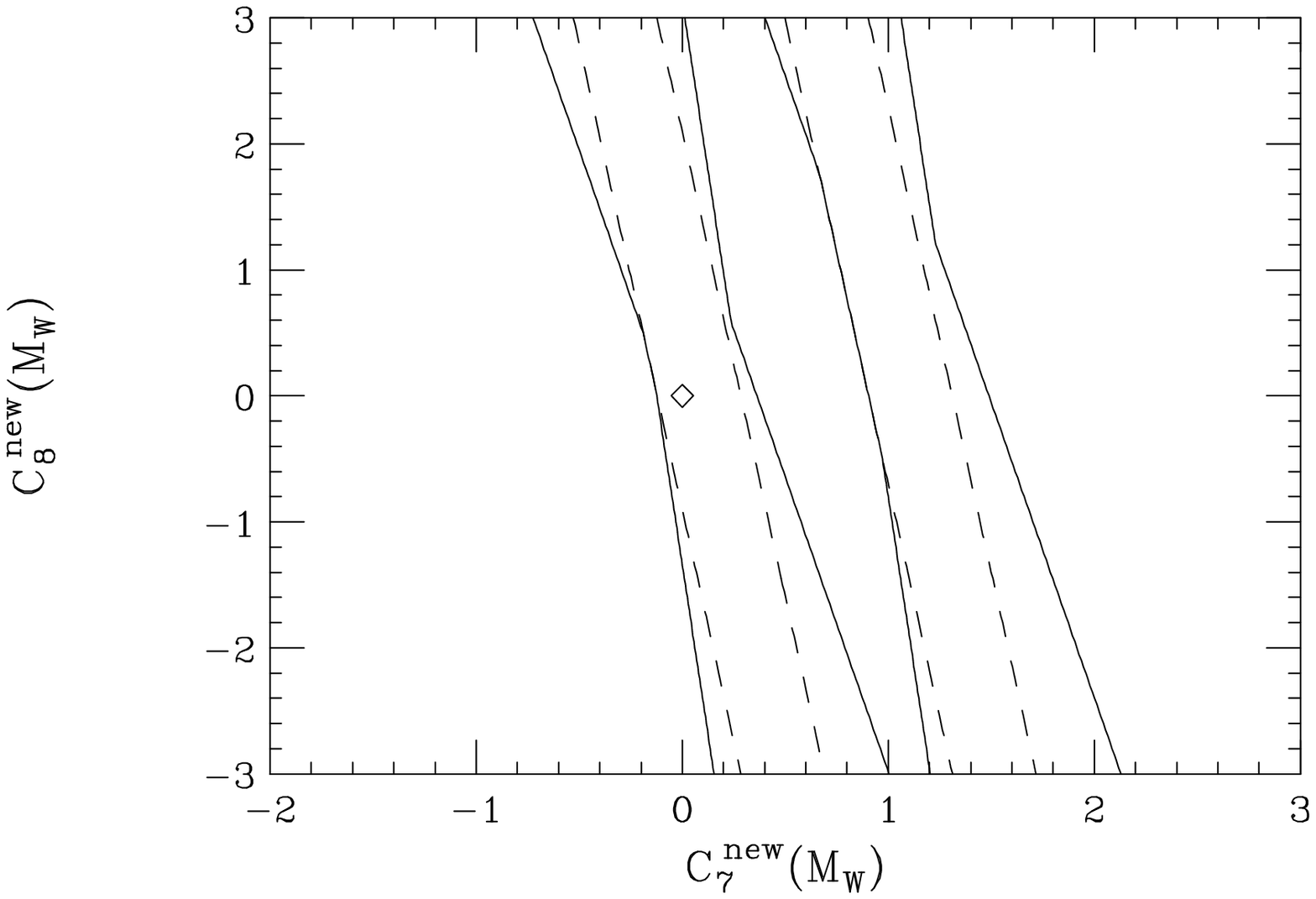}{Bounds
on the contributions from new physics to $C_{7,8}$.  The region allowed
by the CLEO data corresponds to the area inside the solid diagonal bands.
The dashed bands represent the constraints when the renormalization scale
is set to $\mu =m_b$. 
The diamond at the position
(0,0) represents the standard model.}

Measurement of the kinematic distributions associated with the final
state lepton pair in \bsll\ as well as the rate for \bsg\ allows for
the determination of the sign and magnitude of all the Wilson coefficients
for the contributing operators in a model independent
fashion~\cite{ali2,hewett}.  
Here, we perform a Monte Carlo analysis in order to
ascertain how much quantitative information will be obtainable at
future $B$-factories.  We improve upon our previous study\cite{hewett}
by implementing the NLO computations for these decays and by examining
the luminosity dependence of the resulting global fits.  For the
process \bsll, we consider the lepton pair invariant mass distribution
and forward-backward asymmetry for $\ell=e, \mu, \tau$, and the tau
polarization asymmetry for \bstt. We note that the asymmetries have the
form $A(\shat )\sim C_{10} (\hbox{\rm Re}\, 
C_9^{eff} f_1(\shat )+C^{eff}_7 f_2(\shat ))$,
and hence are sensitive probes of the Wilson coefficients. 
We generate ``data,'' assuming the
SM is realized, by dividing the lepton pair invariant mass spectrum
into nine bins.  Six of the bins are taken to be in the low dilepton
invariant mass region below the $J/\psi$ resonance (in order to take
advantage of the larger statistics), with $0.02\leq\shat\leq 0.32$ and
a bin width of $\Delta\shat=0.05$.  We have also cut out the region
near $q^2=0$ in order to remove the photon pole.  The high $M_{\ell^+\ell^-}$
region above the $\psi'$ pole is divided into three bins, corresponding to
$0.6\leq\shat\leq 0.7$, $0.7\leq\shat\leq 0.8$, and $0.8\leq\shat\leq 1.0$.
The number of events per bin is calculated as
\begin{equation}
N_{bin}={\cal L}\int^{\shat_{max}}_{\shat_{min}} {d\Gamma(\bsll)\over d\shat}
d\shat \,,
\end{equation}
and the average value of the asymmetries in each bin is
\begin{equation}
\langle A \rangle_{bin} = {{\cal L}\over N_{bin}}
\int^{\shat_{max}}_{\shat_{min}} A {d\Gamma(\bsll)\over d\shat} d\shat\,.
\end{equation}
We statistically fluctuate the ``data'' using a normalized Gaussian
distributed random number procedure, where the statistical errors are given
by $\delta N=\sqrt N$ and $\delta A=\sqrt{(1-A^2)/N}$.  We expect the errors
in each bin to be statistics dominated and hence we neglect any possible
systematic errors.  For \bsg, we again statistically fluctuate the ``data'' for
the inclusive rate.  However, in this case, the statistical precision will
eclipse the possible systematic and theoretical accuracy.  We thus assume
a flat $10\%$ error in the determination of the branching fraction in
anticipation of future theoretical and experimental improvements.  A three
dimensional $\chi^2$ fit to the coefficients $C_{7,9,10}(\mu)$ is 
performed, employing the usual prescription
\begin{equation} 
\chi^2_i=\sum_{bins} \left(
{Q_i^{obs}-Q_i^{SM}\over\delta Q_i}\right)^2 \,,
\end{equation}
for each observable quantity $Q_i$.  We repeat this procedure for
three values of the integrated luminosity, 
$3\times 10^7$, $10^8$, and $5\times 10^8$ $B\bar B$ pairs,
corresponding to the expected $e^+e^-$ $B$-factory luminosities of one
year at design, one year at an upgraded accelerator, and the total
accumulated luminosity at the end of the programs.  Hadron colliders
will, of course, also contribute to this program, but it is more
difficult to assess their potential systematic and statistical weights
without further study.

The $95\%$ C.L. allowed regions as projected
onto the $C_9(\mu)-C_{10}(\mu)$ and $C^{eff}_7(\mu)-C_{10}(\mu)$ planes are
depicted in Figs.~\ref{fig3}(a-b), where the diamond
represents the expectations in the SM.  We see that the determinations are
relatively poor for $3\times 10^{7}$ $B\bar B$ pairs and that
higher statistics are
required in order to focus on regions centered around the SM.  Clearly,
$C_9$ and $C_{10}$ are highly correlated, whereas $C^{eff}_7$ 
and $C_{10}$ are not.
We see that the sign, as well as the magnitude, of all the coefficients
including $C^{eff}_7$ can now be determined. 

\jfig{fig3}{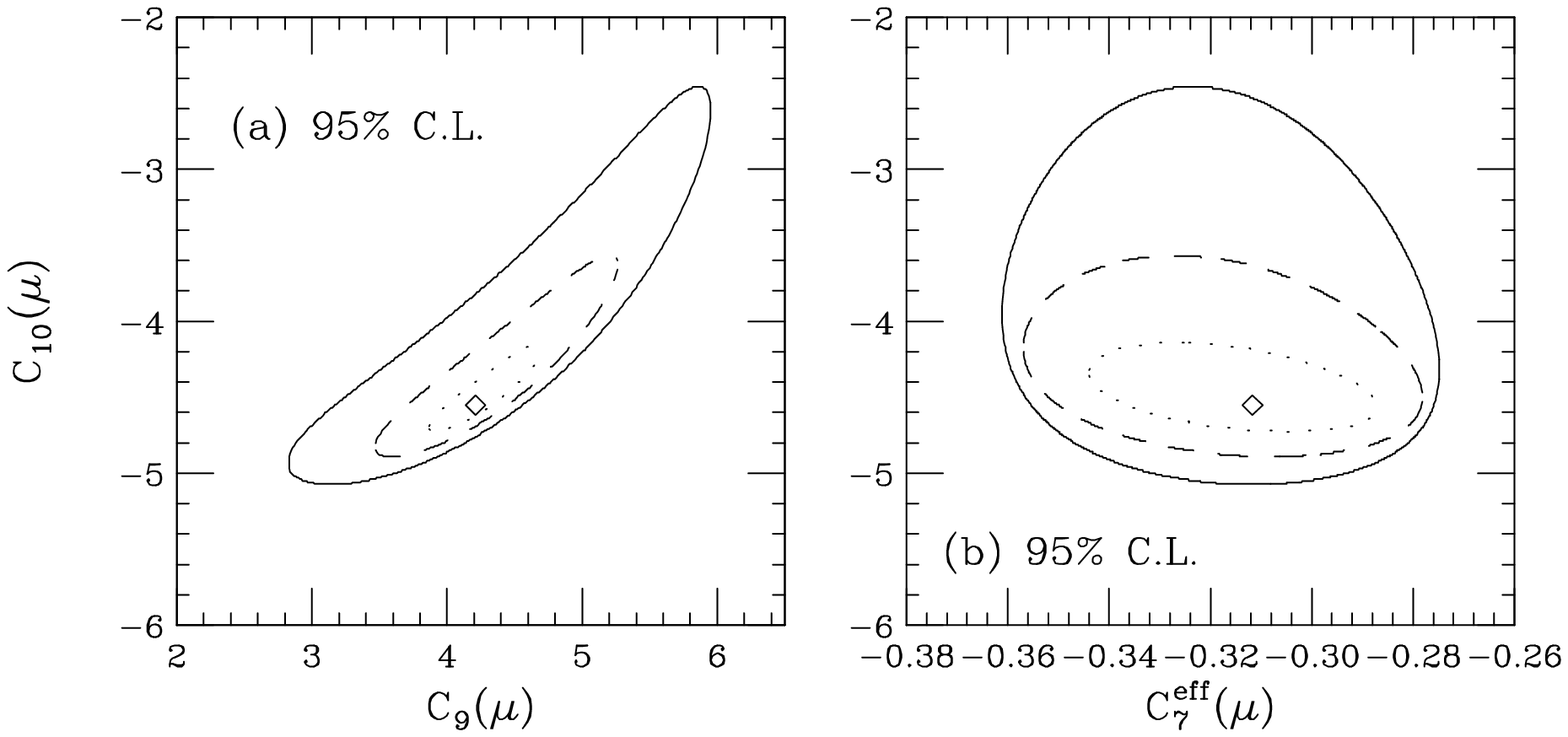}{The $95\%$
C.L. projections in the (a) $C_9 - C_{10}$ and (b) $C^{eff}_7 - C_{10}$ 
planes, 
where the allowed regions lie inside of the contours.  The solid, dashed, and 
dotted contours correspond to $3\times 10^7$, $10^8$, and $5\times 10^8$
$B\bar B$ pairs. The SM prediction is labeled by the diamond.}

For the remainder of this paper, we analyze the supersymmetric contributions
to the Wilson coefficients in terms of the quantities
\beq
R_i\equiv \frac{C^{susy}_i(M_W)}{C^{SM}_i(M_W)}-1\equiv 
{C_i^{new}(M_W)\over C_i^{SM}(M_W)}\,,
\eeq
where $C^{susy}_i(M_W)$ includes the full standard model plus superpartner
contributions.  $R_i$ is meant to indicate a relative fraction difference
from the standard model value.  It is most convenient to define these
ratios at the $W$ scale to avoid the added complication of the renormalization
group evolution to the low scale.

\section{Expectations in supersymmetry}

Supersymmetry has many potential sources for flavor violation.  The
flavor mixing angles among the squarks are {\it a priori} 
separate from the CKM angles of the standard model quarks.  If we
allow the super-CKM angles to be arbitrary then we find, for example,
that the relative SUSY versus SM amplitudes for $b\to s$ are
$(M_W/\tilde m)^n(\tilde V^*_{ts}\tilde V_{tb}/V^*_{ts}V_{tb})$.  
$|V^*_{ts}V_{tb}|\simeq 0.04$, and so
allowing the $\tilde V$ angles to be oriented randomly 
with respect to the CKM angles constitutes
a flavor problem for supersymmetry if $\tilde m$ is
near the weak scale.  
Natural solutions to
this problem exist.  
One solution is alignment~\cite{nir93:337} of the
super-CKM matrices with the quark matrices.  Another natural solution is
universality induced by gauge mediated supersymmetry breaking~\cite{dine}.
In the minimal model of gauge mediated supersymmetry breaking~\cite{dimo1}, 
the $b\to s\gamma$ decay is currently not a strong constraint on the spectrum,
but will show deviations from the standard model at the $B$-factory if
$m_{\tilde \chi^\pm_1}\lsim 350\gev$.
We adopt the viewpoint in this paper that 
flavor-blind (diagonal) soft terms~\cite{dimo2} 
at the high scale are the phenomenological
source for the soft scalar masses at the high scale, and that the CKM
angles are the only relevant flavor violating sources.  We build on
other studies of supersymmetry effects on rare $B$ 
decays~\cite{rareb,ali2,cho96:360}.

The spectroscopy of the supersymmetric states is model dependent.  We will
analyze two possibilities.  The first possibility is that all the
supersymmetric states follow from common scalar mass at high scale
and common gaugino mass at the high scale.  This is the familiar
minimal supergravity model.  The second possibility is to 
relax the condition of common scalar masses at the high scale and allow
them to take on more uncorrelated values at the low scale
while still preserving gauge invariance.

We begin by searching over the full parameter space of minimal supergravity
model.  We use the words ``minimal supergravity'' as an idiom to indicate that
we generate~\cite{kkrw} these models by applying common 
soft scalar masses
and common gaugino masses at the boundary scale.  The tri-scalar $A$ terms
are also an input at the high scale and are universal.
The radiative electroweak symmetry 
breaking conditions yield the $B$
and $\mu^2$ terms as output, with a $\mbox{sign}(\mu )$ ambiguity left over 
as an input parameter. (Here $\mu$ refers to the Higgsino mixing
parameter.) We also choose $\tan\beta$ and restrict it
to a range which will yield perturbative Yukawa couplings up to the GUT scale.

We have generated thousands of solution according to the above procedure.  The
ranges of our input parameters are $0< m_0 < 500\gev$, $50< m_{1/2} < 250\gev$,
$-3 < A_0/m_0 < 3$, $2 < \tan\beta < 50$,
and we have taken $m^{phys}_t=175\gev$.  Each supersymmetric solution is
kept only if it is not in violation with present constraints from
SLC/LEP and Tevatron direct sparticle production limits.  
For each of these remaining solutions 
we now calculate $R_{7-10}$~\cite{anlauf94:245}.   
Expansions of
$C_i(\mw )$ with the assumption of approximate 
universality are given in the appendix.  

First, we present a scatter plot of
$R_7$ vs. $R_8$  in Fig.~\ref{fig4}; we remind the reader that these quantities
are evaluated at the electroweak scale.
Again, each point in the scatter plot is derived from the minimal supergravity
model with different initial conditions. Also, each point is consistent with
all collider bounds and is out of reach of LEPII.  The first thing to note
from the figure is that large values of $R_7$ and $R_8$
are generated, and the $R_7$ and $R_8$ values are very strongly correlated.
The diagonal
bands represent the bounds on the Wilson coefficients from the observation of
\bsg\ as determined in the 
previous section.  
We note that these bands appear to be straight here as they correspond to a 
small region of Fig.~\ref{fig2}.
We see that the current CLEO data already places 
significant restrictions on the supersymmetric parameter space.  Further
constraints will be obtainable once a $10\%$ measurement
of $B(B\to X_s\gamma )$ is made, and the sign of $C_7$ is determined from
a global fit described in the previous section.
In this case, if no
deviations from the SM are observed, the supersymmetric contributions will 
be restricted to lie in the dashed band.  It is clear that these
processes can explore vast regions of the supersymmetric parameter space.
In fact, it is possible that spectacularly large deviations in
rare $B$ decays could be manifest at $B$ factories, while collider experiments
would not detect a hint of new physics.

\jfig{fig4}{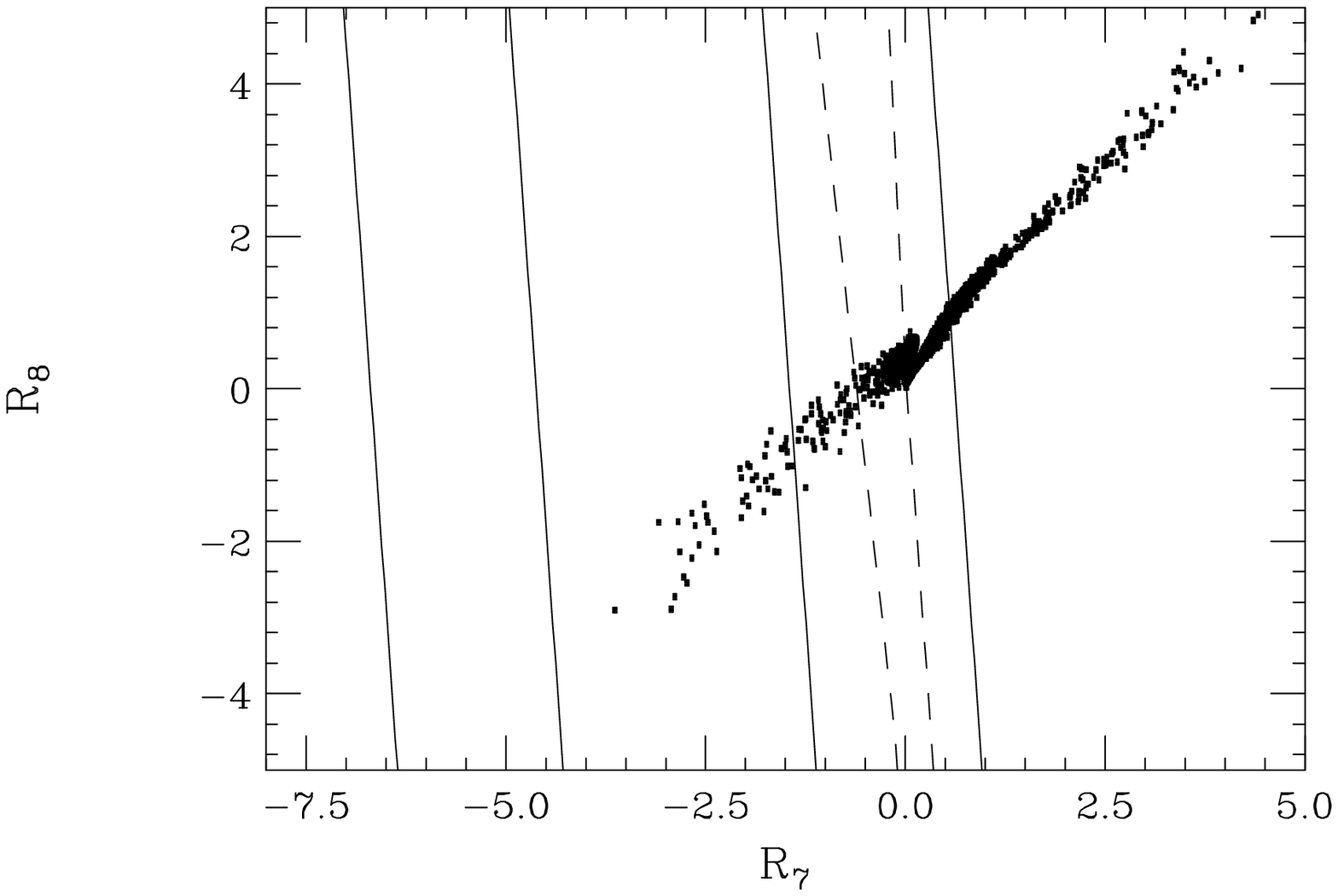}{Parameter 
space scatter plot of $R_7$ vs. $R_8$ in minimal supergravity model.
The allowed region from CLEO data, as obtained in Fig.~2, lies inside the
2 sets of solid diagonal bands.  The dashed band represents the potential
$10\%$ measurement from the previously described global fit to the
coefficients.}

The large effects in $R_7$ and $R_8$ are coming from
models with $|\mu |\lsim 400\gev$ as can be seen in Fig.~\ref{fig5}.
This is because light charged Higgsinos, or rather light charginos with a
large Higgsino fraction, are required in order to yield a large effect
on the Wilson coefficients.  Later in this section we will
demonstrate this requirement more carefully
by expanding the supersymmetric contributions
in the Higgsino limit.  

\jfig{fig5}{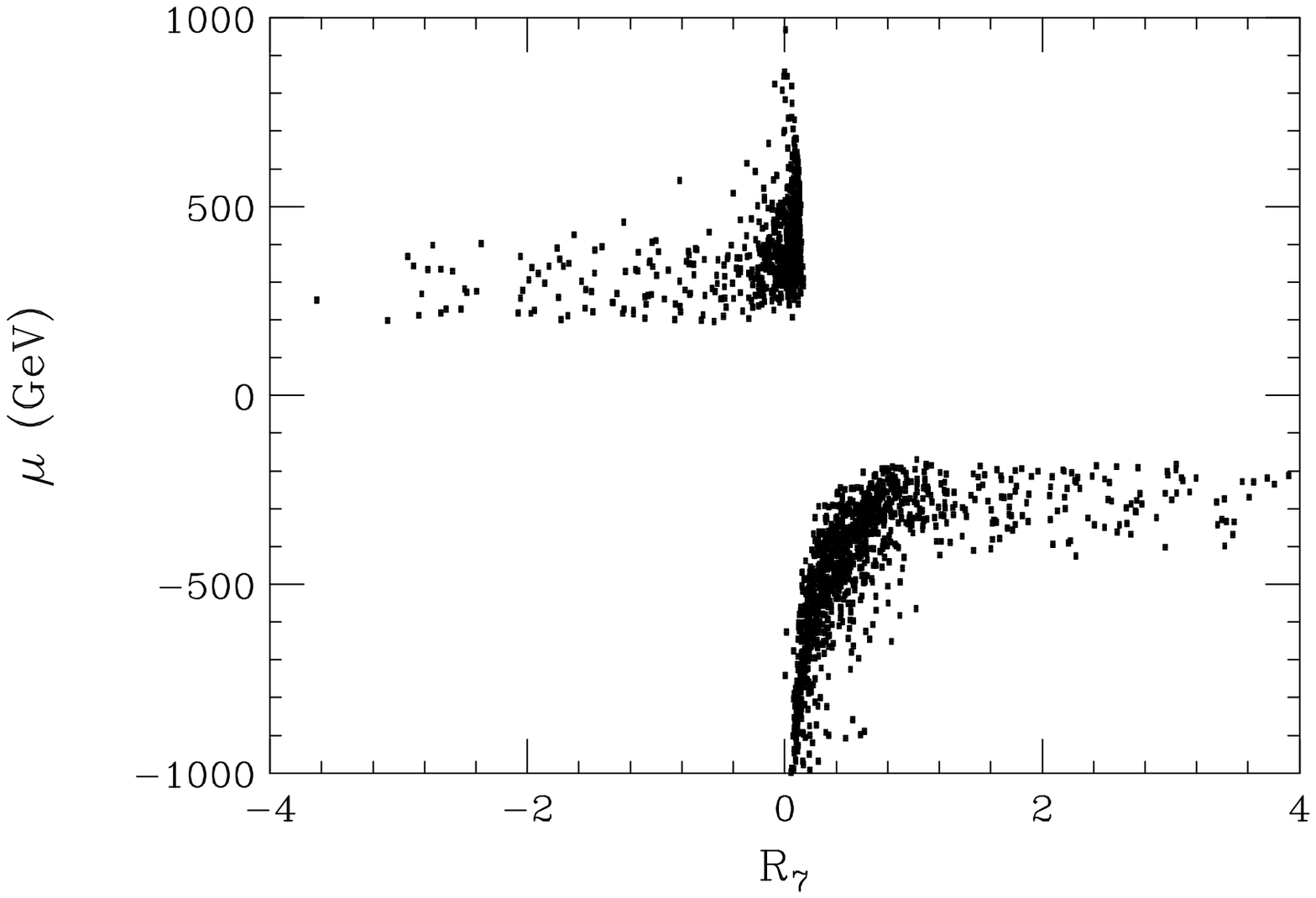}{Parameter 
space scatter plot of $R_7$ vs. $\mu$ in minimal supergravity model.}

In Fig.~\ref{fig6} the correlation
between $R_9$ and $R_{10}$ is plotted using the same supersymmetric
parameter space.  We see that $R_9$ is always positive since charged
Higgs and chargino contributions always add constructively.
\jfig{fig6}{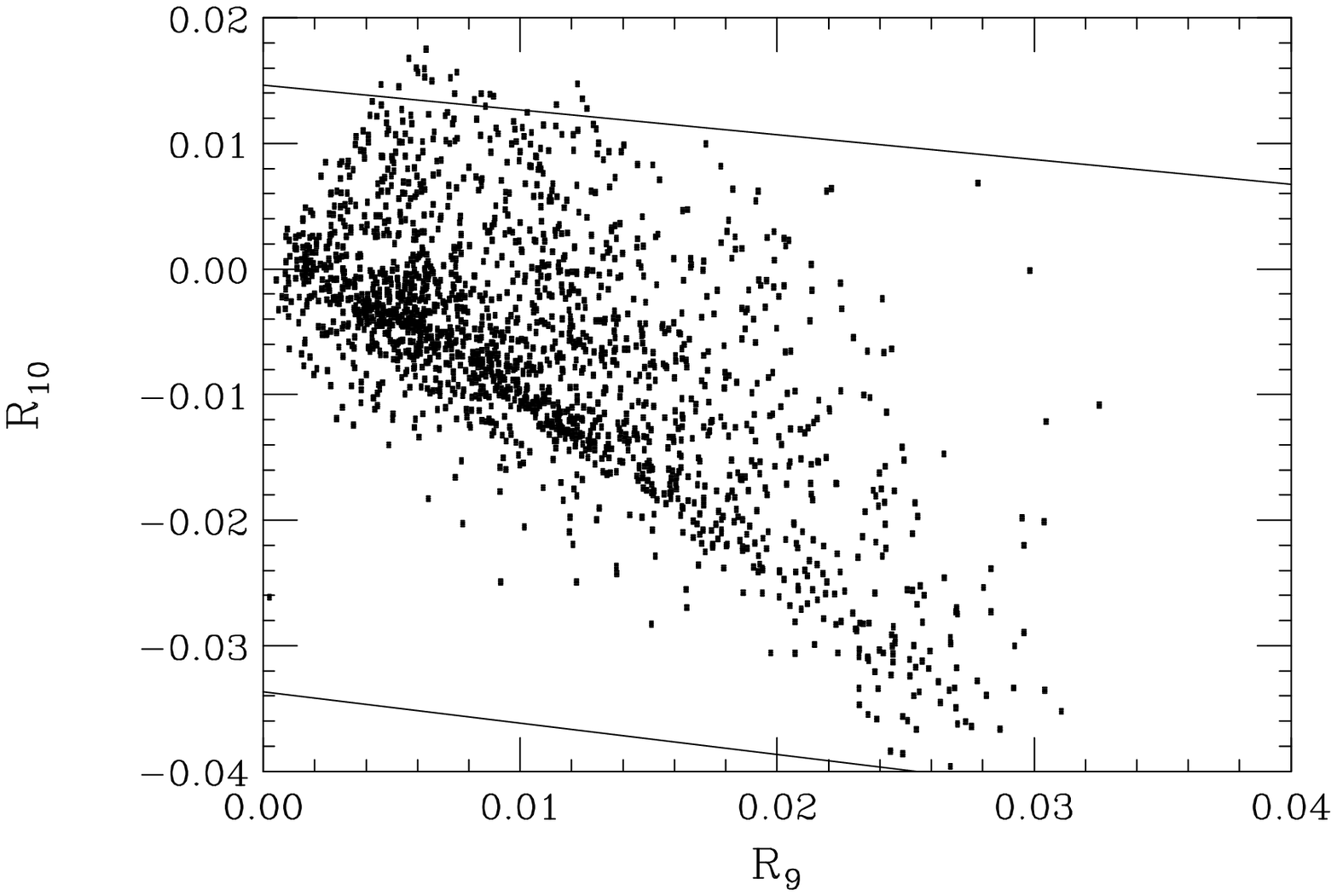}{Parameter 
space scatter plot of $R_9$ 
vs. $R_{10}$ in minimal supergravity model. The global fit to the coefficients
obtained in Fig.~3 with $5\times 10^8\, B\bar B$ pairs
corresponds to the region inside the diagonal bands.}
We see that the values of $R_9$ and $R_{10}$ are bounded by
about $0.04$, a small number compared to the range for $R_7$.  
The main reason for
these smaller values is the dependence on the bottom Yukawa 
$\lambda_b\propto 1/\cos\beta$.  $R_7$ has a contribution directly
dependent on this $1/\cos\beta$ Yukawa enhancement, 
and the other multiplicative
terms associated with this Yukawa are the large top Yukawa and a
large kinematic loop factor.  $R_{9}$ and $R_{10}$ do not have such factors
due the chirality structure of these operators and the
requirement that leptons and sleptons only couple via $SU(2)$ and $U(1)$
gauge couplings.
These factors,
along with the correlations between the mass spectra dictated by
minimal supergravity relations, render the minimal supergravity 
contributions to
$R_{9,10}$ essentially unobservable.  The solid lines in this figure
correspond to the 95\% C.L. bounds obtainable with high integrated luminosity
($5\times 10^8$ $B\bar B$ pairs) at $B$ factories 
from the global fit performed in the previous section.
If large deviations in $R_{9,10}$ are observed, then,
of course, that would be an indication that the minimal model presently
under discussion is not the correct description of nature.
Later in this section we will find that even when the mass correlations
of the minimal supergravity model are lifted, the quantities $R_{9,10}$ 
still cannot be large. 

We next examine the effects of the minimal supergravity model
on the kinematic distributions for \bsll.  Using our generated models 
as input, we calculate the maximal deviations 
from the SM for the $M_{\ell^+\ell^-}$ distribution, lepton pair 
forward-backward asymmetry, and tau polarization asymmetry.
These are displayed in Figs.~\ref{fig7}(a-c).  Here, the dotted line
corresponds to the SM prediction, while the  dashed (solid) bands
represent the maximal possible deviations due to points in
the supersymmetric parameter space which are within (outside) the expected
reach of LEPII.  We have also demanded consistency with the present 
CLEO data on
\bsg. We see from the figure that, generally, larger deviations are expected
for the asymmetries than for the branching fraction, and that constraints
from LEPII on supersymmetry greatly affect the magnitude of these potential
deviations. Since the SUSY contributions to $R_{9,10}$ are so small
these deviations are mainly due to $R_{7,8}$.
We find that much larger effects in these distributions 
are possible if
the constraint from radiative $B$ decays is not taken into account.

\jfig{fig7}{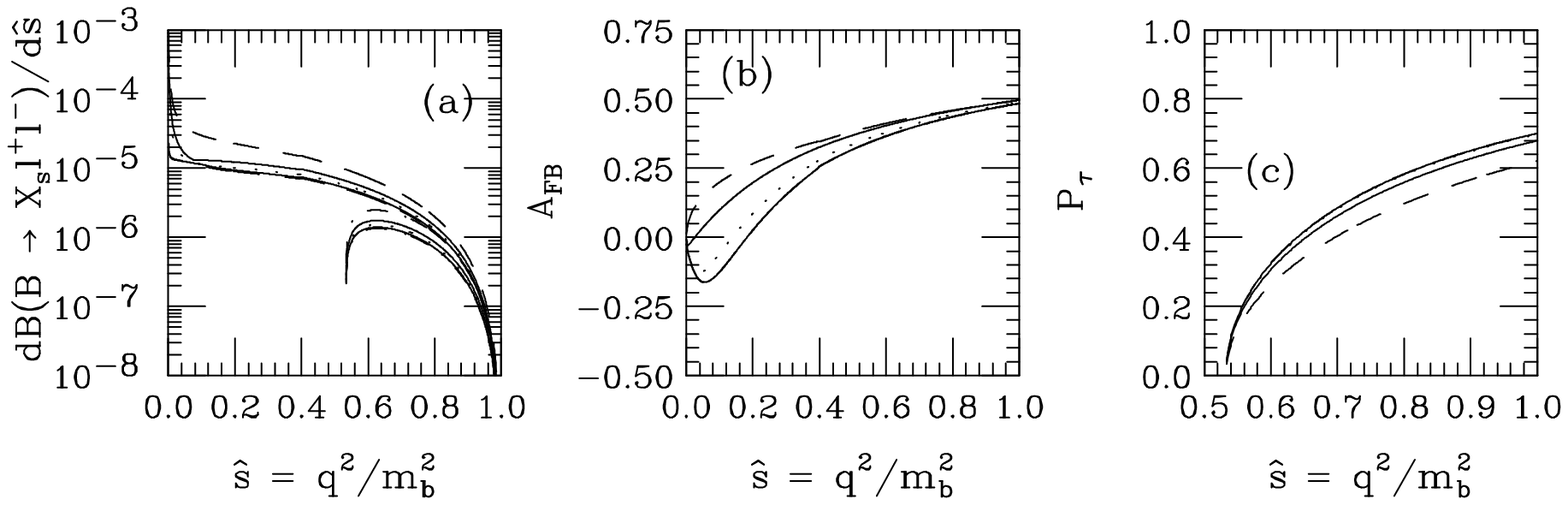}{The (a) differential branching fraction,
(b) lepton pair forward-backward asymmetry, and (c) tau polarization
asymmetry as a function of the scaled momentum transfer to the lepton
pair, $\hat s\equiv q^2/m_b^2$.  The dotted curves represent the SM
prediction, while the dashed and solid bands correspond to the
maximal potential deviations due to supersymmetric contributions for
different regions of the parameter space as described in the text.
In some cases the dashed line overlaps with the solid line.}

We now adopt a more phenomenological approach.
The maximal effects for the parameters $R_i$ can be estimated
for a superparticle spectrum independent of these high scale assumptions.
However, we still maintain the assumption that CKM angles alone constitute
the sole source of flavor violations in the full supersymmetric lagrangian.
We will focus on the region 
$\tan\beta \lsim 30$ since enormous effects are possible
in the large $\tan\beta$ limit; later on we will discuss the large 
$\tan\beta$ limit more
carefully.  The most important features which result in large effects are
a light $\tilde t_1$ state present in the spectrum and at least one
light chargino state.  For the dipole moment operators a light Higgsino
is most important. 
A pure higgsino and/or pure
gaugino state have less of an effect than two mixed states when searching
for maximal effects in $C_9$ and $C_{10}$.  
In fact, we have found that $M_2 \simeq 2 \mu$ is optimal.

Fig.~\ref{fig8}, and \ref{fig9} display the
maximum contribution to $R_{9,10}=C^{susy}_{9,10}(M_W)/C^{SM}_{9,10}(M_W)-1$ 
versus an applicable SUSY mass scale.  
\jfig{fig8}{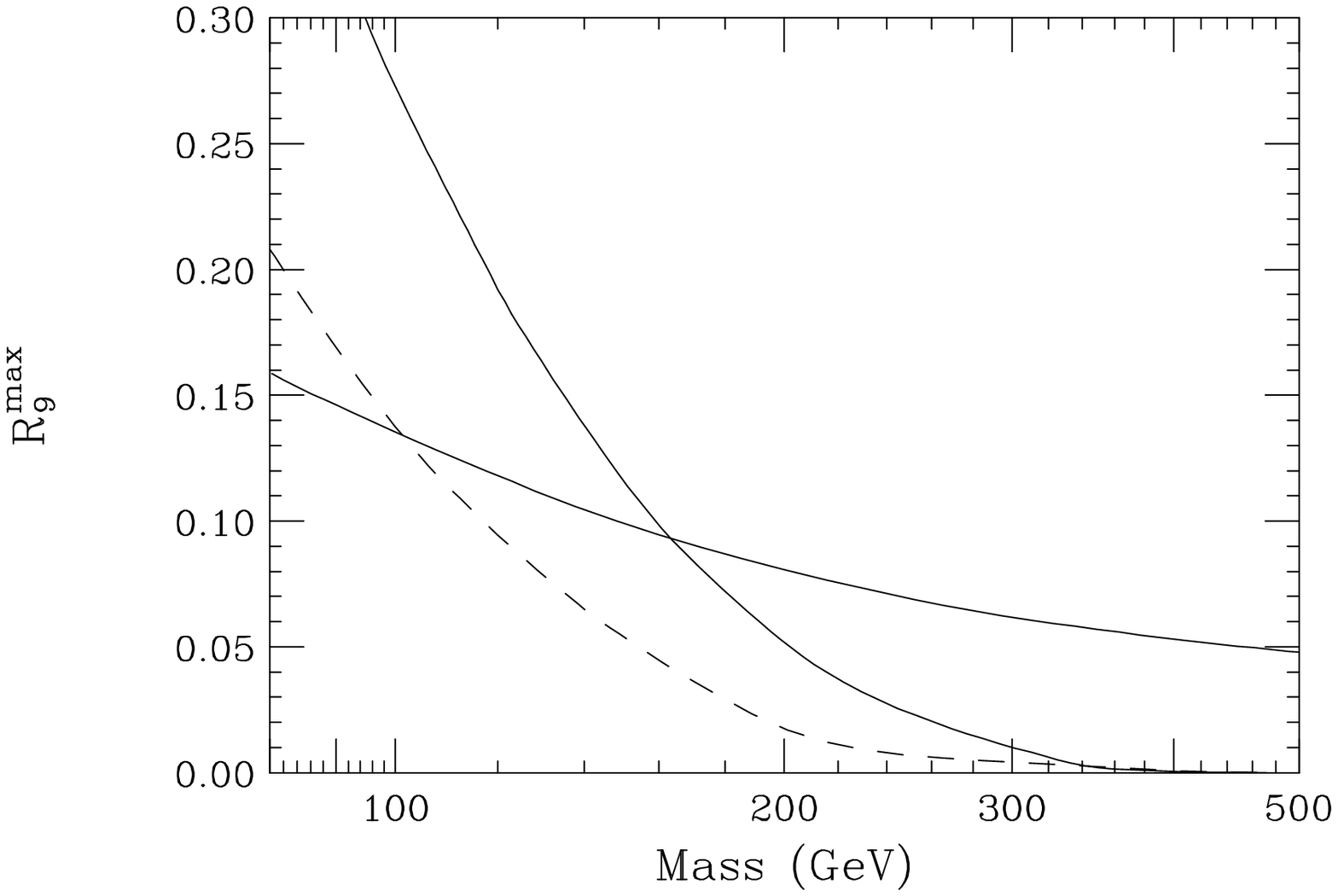}{
The maximum value of $R_9=C^{susy}_9(M_W)/C^{SM}_9(M_W)-1$ 
achievable for general 
supersymmetric models. The top solid line comes from $t-H^\pm$
contribution and is displayed versus the $H^\pm$ mass.  The bottom solid line
is from $\tilde t_i -\chi^\pm_j$ contribution with $\tan\beta =1$
and is shown versus the $\chi^\pm_i$ mass.  The dashed line is
the $\tilde t_i -\chi^\pm_j$ contribution with $\tan\beta =2$.
The other mass parameters which are not plotted are chosen to be just above
LEPII and Tevatron's reach.}
\jfig{fig9}{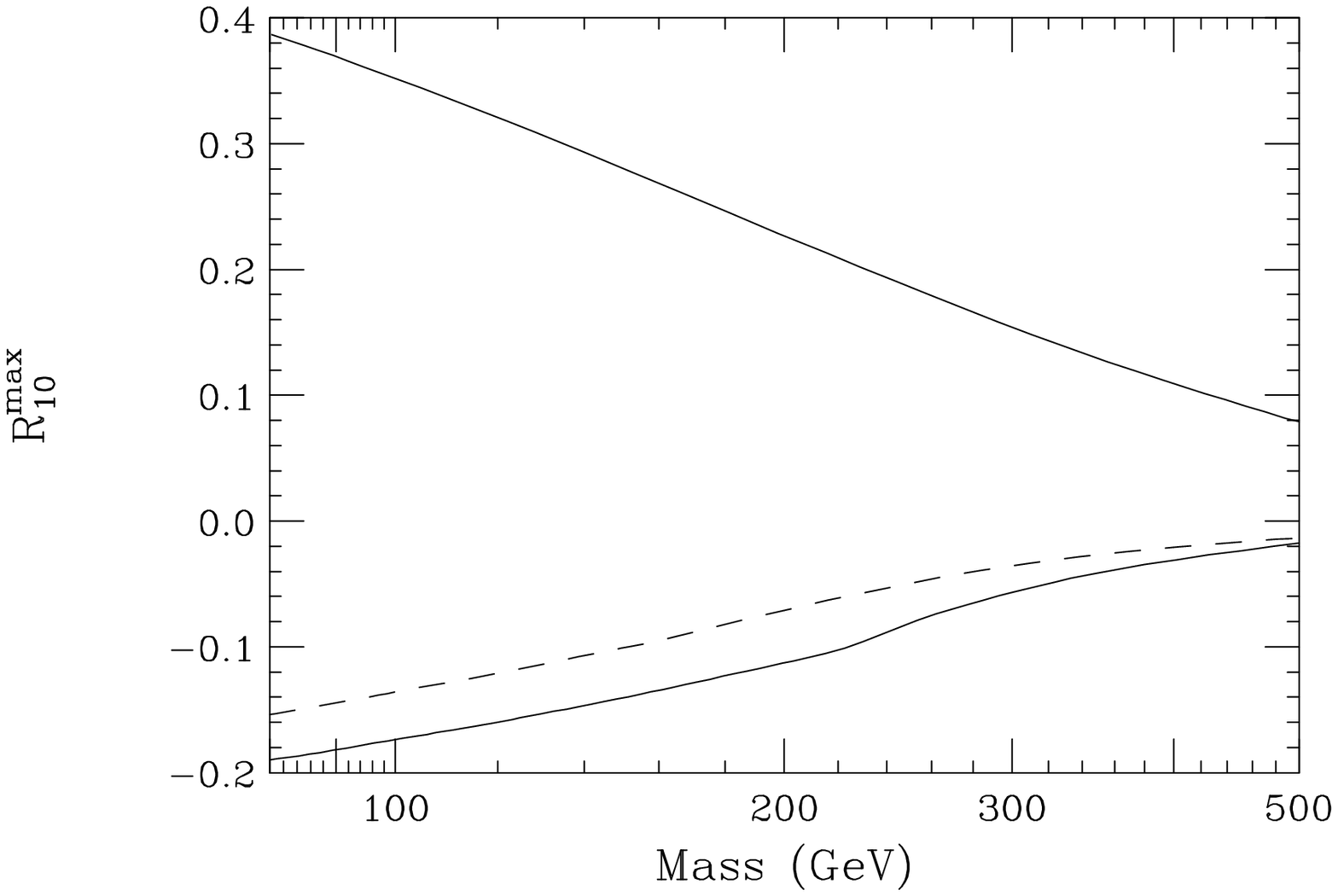}{
The maximum value of $R_{10}=C^{susy}_{10}(M_W)/C^{SM}_{10}(M_W)-1$ 
achievable for general 
supersymmetric models. The top solid line comes from $t-H^\pm$
contribution and is shown versus the $H^\pm$ mass.  The bottom solid line
is from $\tilde t_i -\chi^\pm_j$ contribution with $\tan\beta =1$
and is plotted versus the $\chi^\pm_i$ mass.  The dashed line is
the $\tilde t_i -\chi^\pm_j$ contribution with $\tan\beta =2$.
The other mass parameters which are not presented are chosen to be just above
LEPII and Tevatron's reach.}
The other masses which are not plotted ($\tilde t_i$,
$\tilde l_L$, etc.) are chosen to be just above the reach of LEPII or
the Tevatron, whichever gives better bounds.  

The maximum size of $R_{9,10}$ is much larger than what was allowed in 
the minimal supergravity model. The reason for this is the lifted restriction
on mass correlations.  Light
sleptons, sneutrinos, charginos, and stops are allowed simultaneously
with mixing angles giving 
the maximal contribution to the $R_i$'s.  However, we find that the
maximum allowed values for $R_{9,10}$ are still much less than unity.  
Earlier we determined that $B$ factory data would be sensitive to
$\Delta R_9\gsim 0.3$ and $\Delta R_{10}\gsim 0.08$ at the highest 
luminosities, 
and so the largest SUSY effect would give a
$~1-2\sigma$ signal in $R_{9,10}$, hardly enough to be a compelling indication
of physics beyond the standard model.  If, on the other hand, much larger
deviations of $R_{9,10}$ are found in the data, it could mean 
the assumption of only CKM angles allowed in the supersymmetric
lagrangian is inaccurate, or it could  indicate that minimal 
supersymmetry is not the source of physics beyond the standard model. 

It should be remembered that even though it  
appears difficult to resolve the SUSY
contributions to the coefficients $C_{9-10}$, the
$B\to X_sl^+l^-$ decay rate and distributions 
can still demonstrate large deviations from the standard model
induced largely by the SUSY
corrections to $C_7(M_W)$.  The global fit using all the rare $B$ decay data
is still necessary in this circumstance
since it will enable a precise determination in which band in the
$C_7$ vs.~$C_8$ plot we reside.  Furthermore, some ideas~\cite{nelson} 
of physics beyond
the standard model predict small corrections to $B\to X_s\gamma$ and
large deviations in $B\to X_sl^+l^-$, motivating again the procedure of
doing a global fit to all the rare $B$ decay data.

Given the sensitivity of all the 
observables it is instructive to narrow the focus to $C_7(M_W)$.  
In the minimal
supergravity models, the scalars obtain dependence on the gaugino masses
through the renormalization group running.  However, the gaugino masses
do not get substantial scalar mass contributions to their masses.  This
tends to separate the mass scale for the scalars far from the gauginos.  
The separation is especially important between electroweak gauginos and
strongly interacting squarks.  Neglecting the D-term contributions the
squarks have masses given roughly by
$\tilde m^2_q  \approx m_0^2 +7 m^2_{1/2}$
and the weak gaugino has mass
$m_{\tilde W} \approx 0.8 m_{1/2}$
where $m_0$ is the common scalar mass and $m_{1/2}$ is the common
gaugino mass at the high scale.  From these equations it is easy to
see that the squark masses are much heavier than the weak gaugino mass
for any given $m_{1/2}$ and $m_0$.

When all the squark masses are very heavy (much heavier than the
charginos, for example) then the SUSY contributions to $C_7(M_W)$ decouple.
However, one eigenvalue of the stop squark mass matrix might be much
lighter than the other squarks.  The large top Yukawa tends to drive
$\tilde t_R$ much lower than the other $\tilde q_{L,R}$.  Also,
the stop squark has a mixing term
proportional to $m_t (A_t-\mu\cot\beta)$.  Since this mixing is proportional
to the top mass, it can be substantial.  Mixings in any positive definite
matrix will push the lightest eigenvalue lower and the heaviest eigenvalue
higher.  These two effects tend to push
the lightest
stop eigenvalue well below the other squarks.  In fact, a highly mixed,
light stop squark is generic in these theories.  For a large supersymmetric
contribution to $R_b$~\cite{Rbpapers} and/or the ability to achieve 
successful baryogenesis in the early universe~\cite{transition}, 
a light $\tilde t_R$ is needed.
We would therefore like to present results on $C_7(M_W)$ in the limit of
one light squark, namely the $\tilde t_1$, and light charginos.  
We allow the $\tilde t_1$ to have
arbitrary components of $\tilde t_L$ and $\tilde t_R$ since cross terms
can become very important.  This is especially noteworthy 
in the high $\tan\beta$ limit as we will discuss below.

In this limit of one light top squark, we can expand the
chargino-stop contribution to $C_7(M_W)$ as
\beq
\label{mainc7}
\delta C_7 (M_W) =\frac{1}{6} \sum_i \frac{M_W^2}{m_{\chi^\pm_i}^2}
\lambda^L_i \left\{ \lambda^L_i F(x_i)
+2\lambda^R_i\frac{m_{\chi^\pm_i}}{m_b}
J(x_i)\right\}\,,
\eeq
where
\bea
\lambda^L_i & = & -T_{11} V_{i1}+T_{12}
      \frac{V_{i2}}{\sqrt{2}} \frac{m_t}{M_W}\frac{1}{\sin\beta}\,,\nonumber \\
\lambda^R_i &=& T_{11} 
\frac{U_{i2}}{\sqrt{2}}\frac{m_b}{M_W}\frac{1}{\cos\beta}\,,\nonumber \\
x_i & = & \frac{m^2_{\tilde t_1}}{m^2_{\chi^\pm_i}}\,, \\
J(x) & = & \frac{5-7x}{2(1-x)^2}+\frac{2x-3x^2}{(1-x)^3}\log x\,,\nonumber \\
F(x) &= & \frac{7-5x-8x^2}{6(1-x)^3}+\frac{2x-3x^2}{(1-x)^4}\log x\,.\nonumber
\eea
The matrix $T_{ij}$ is the stop mixing matrix which rotates 
$(\tilde t_L \tilde t_R)$ into $(\tilde t_1 \tilde t_2)$.  The matrices
$U_{ij}$ and $V_{ij}$ are the usual chargino mixing matrices~\cite{hk}.
For the reader's convenience we have plotted the functions $F(x)$ and
$J(x)$ in Fig.~\ref{fig10}.

\jfig{fig10}{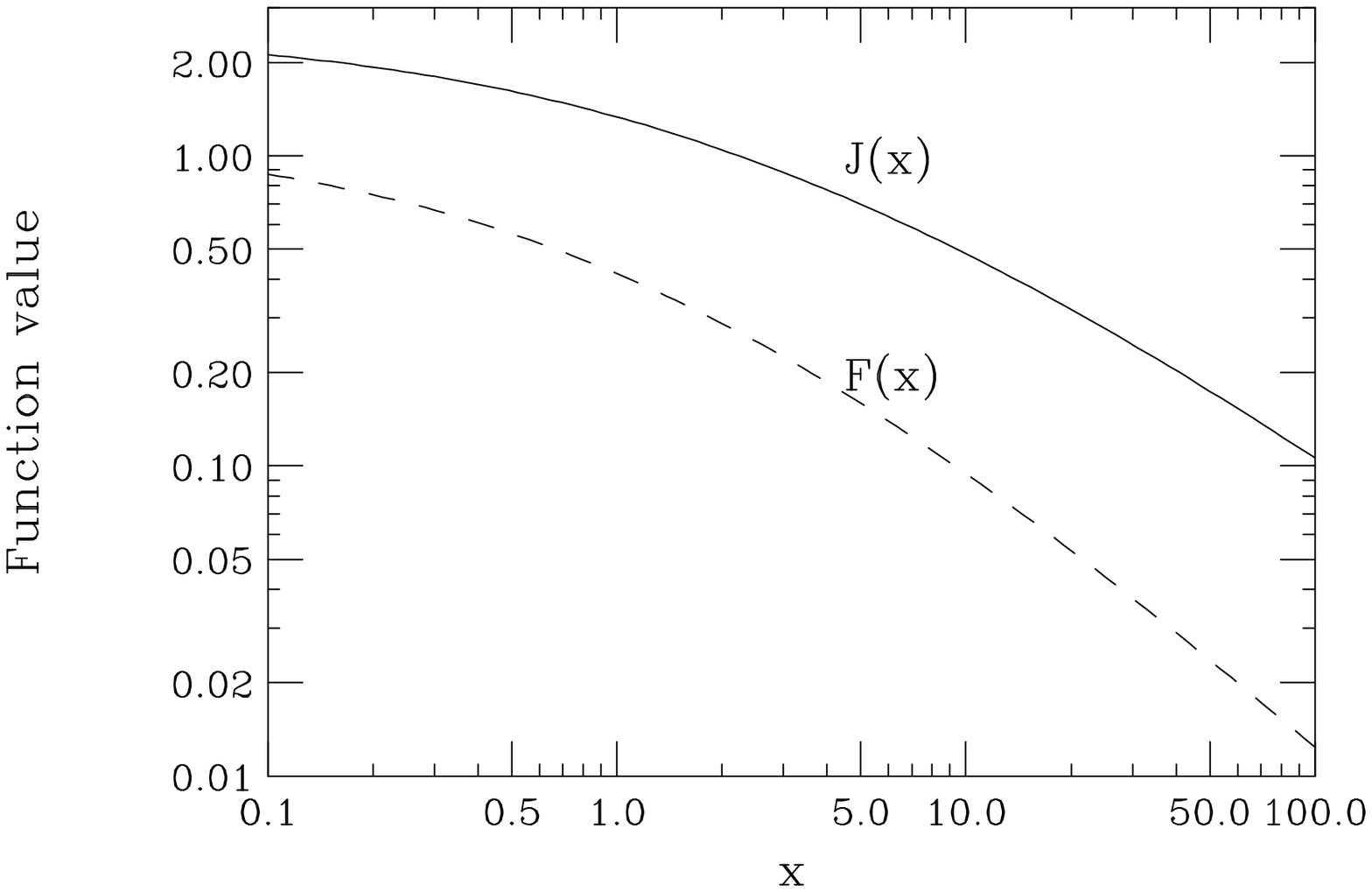}{
The kinematic loop functions $F(x)$ and $J(x)$ necessary to
calculate the standard model and supersymmetric contributions
to $C_7(M_W)$.  $J(x)$ and $F(x)$ asymptote to $5/2$ and $7/6$ respectively
as $x\to 0$.}

The total contribution to $\delta C_7(\mw)$ above
will depend on several combinations of
mixing angles: $T_{11}^2$, $T_{11}T_{12}$, $V_{11} U_{12}$, etc, and
cancellations can occur for different signs of $\mu$~\cite{garisto93:372}.  
Therefore, it is instructive to summarize the
relative signs of each angle combination in the evaluation of
Eq.~\ref{mainc7}:
\bea
\sgn (U_{12}V_{11}) & = & -\sgn (\mu )\,,\nonumber \\
\sgn (U_{12}V_{12}) & = & +\sgn (\mu )\,,\nonumber \\
                    &   & +,~\mbox{if}~M_2\tan\beta +\mu < 0 \,,\nonumber \\
\sgn (V_{11}V_{12}) & = & -\sgn (M_2\tan\beta +\mu )\,, \\
\sgn (T_{11}T_{12}) & = & +,~\mbox{if}~\tilde t_R<\tilde t_L
                           ~\mbox{and}~A_t-\mu\cot\beta < 0 \,,\nonumber\\
                    &   & -,~\mbox{if}~\tilde t_R<\tilde t_L
                           ~\mbox{and}~A_t-\mu\cot\beta > 0\,, \nonumber \\ 
                    &   & -,~\mbox{if}~\tilde t_L<\tilde t_R
                           ~\mbox{and}~A_t-\mu\cot\beta < 0 \,,\nonumber \\
                    &   & +,~\mbox{if}~\tilde t_L<\tilde t_R
                           ~\mbox{and}~A_t-\mu\cot\beta > 0  \,.\nonumber
\eea
We are using the Haber-Kane convention for the sign of $\mu$ which requires,
for example, that $+\mu$ be in the chargino mass matrix.

The first case we discuss is the limit where the lightest chargino is
a pure Higgsino and the lightest stop is pure right-handed:  
$\ca \sim \tilde H^\pm$, $\tilde t_1 \sim \tilde t_R$.  Then Eq.~\ref{mainc7}
can be written as
\beq
\delta C_7 (M_W) = \frac{1}{12}\frac{m^2_t}{m^2_{\ca}}
F\left( \frac{m^2_{\tilde t_1}}{m^2_{\ca}}\right)\,.
\eeq
This limit is roughly applicable in the case where $R_b$ is affected
substantially by supersymmetric corrections~\cite{Rbpapers,brignole96:293}.  
After LEP~II completes its run the charginos will
have been probed up to about $M_W$, and so the maximum effect 
on $\delta C_7(M_W)$
that would be possible in this limit (if LEP~II does not find charginos)
corresponds to setting $m_{\ca} \simeq M_W$.   We do this and show the result
as a function of the $\tilde t_R$ mass in Fig.~\ref{fig11} (dashed line).  
The contribution to $C_7(M_W)$ in this limit is always positive.  Since 
$C^{SM}_7(M_W)$ is a negative quantity in our convention, then 
$R_7=\delta C_7(M_W)/C^{SM}_7(M_W)$ is necessarily negative as well.

\jfig{fig11}{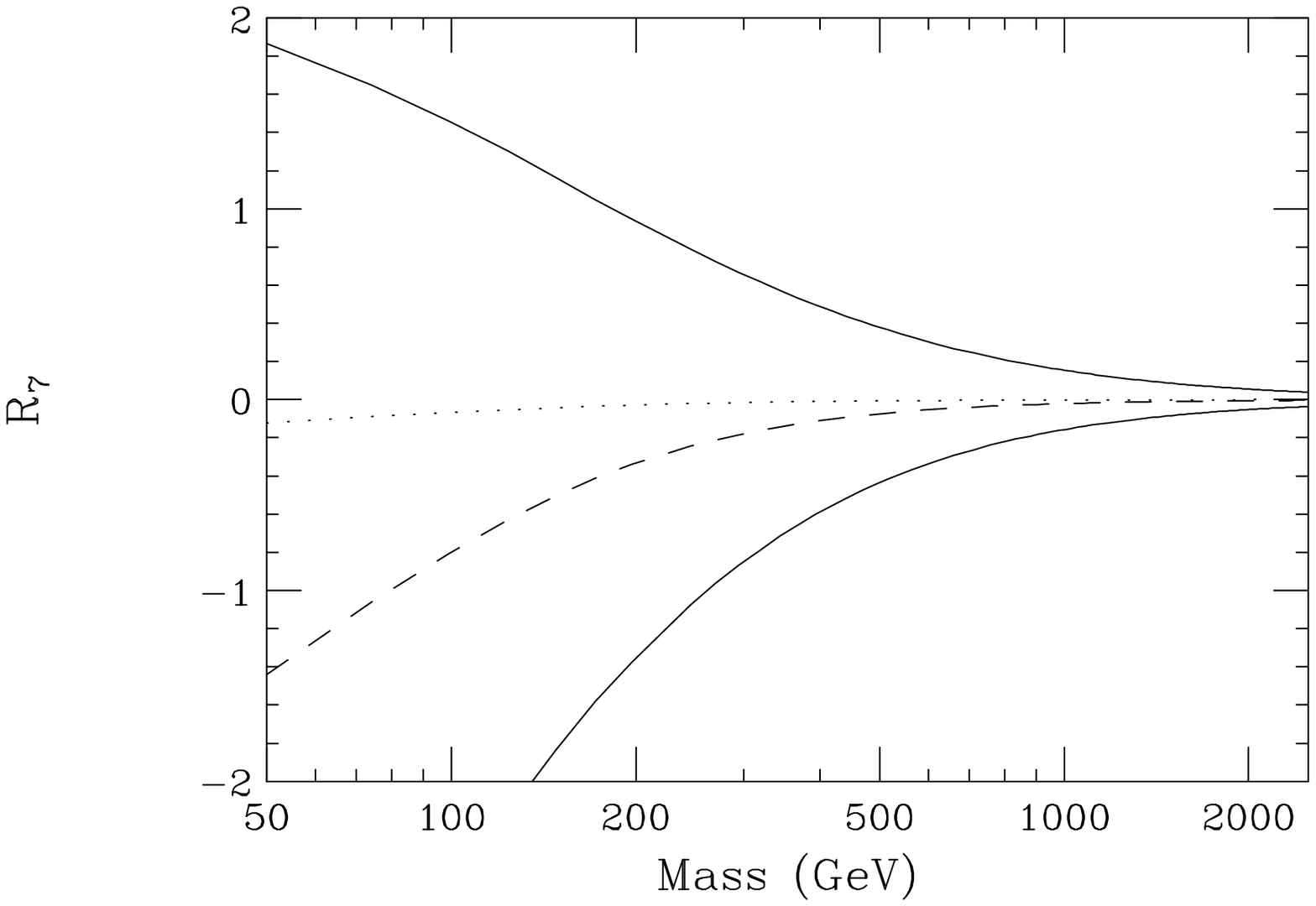}{Contributions 
to $R_7$ in different limits.  The top solid line
is the charged $H^\pm/t$ contribution versus $m_{H^\pm}$.  
The bottom solid line is the
$\tilde \chi^\pm_1/\tilde t_1$ contribution versus $m_{\tilde \chi^\pm}$
where both 
the chargino and stop are maximally
mixed states (equal magnitude mixtures for all states).
This line was made for $\mu <0$.   
The dashed line is the $\tilde H^\pm /\tilde t_R$ contribution,
and the dotted line is the $\tilde W^\pm /\tilde t_1$ contribution 
($\tilde t_1\propto \tilde t_R +\tilde t_L$ is a maximally mixed stop
mass eigenstate).  These two lines are both plotted against the
$\tilde \chi^\pm_1$ mass.
All lines are for $\tan\beta =2$ and $m_t=175\gev$.
We have set all other masses to be just above the reach of LEPII.}

In the case where the only light chargino is a pure Wino we find,
\beq
\delta C_7 (M_W) = \frac{T_{11}^2}{6}\frac{m^2_W}{m^2_{\tilde W}}
F\left( \frac{m^2_{\tilde t_1}}{m^2_{\tilde W}}\right) \,.
\eeq
If $\mu$ is very large and $\tan\beta$ is not too large, then this contribution
will be the largest to $\delta C_7(\mw)$.  
The effects of a light pure Wino are generally
small since the Wino couples like $g_2$ rather than the top Yukawa, and since
generally supersymmetric models do not yield light $\tilde t_L$ necessary
to couple with the Wino.  The loop integral $F(x)$ which characterizes the
pure gaugino contribution is also small.
Therefore, these contributions are both coupling
suppressed and ``loop integral'' suppressed.  
The contribution of a pure
Wino to $R_7$ is shown in Fig.~\ref{fig11} (dotted line).  As expected, this
contribution is rather small and always negative.

As we pointed out above, in minimal supergravity models, what we mostly expect
is a highly mixed $\tilde t_1$ state such that all entries in the $T_{ij}$
are filled.  By just looking at the limit of pure $\tilde t_R$ some 
very interesting
effects can be missed.  For example, an attractive high scale theory is
supersymmetric $SO(10)$.  These have been shown to successfully recover many
important features of the standard model:  
quark to lepton ratios, CKM angles, etc.
One of the generic predictions in these models is that $\tan\beta$ must be
rather high ($\sim m_t/m_b$) in order to get the $b-\tau -t$ mass unification.
Therefore, it is interesting to focus on contributions to $C_7(M_W)$ which are 
especially dependent on high $\tan\beta$.  Since 
$\lambda^R_i \propto 1/\cos\beta \sim \tan\beta$, we isolate this piece.  
For simplicity we
expand about $\ca \sim \tilde H^\pm$, although it is clear that 
substantial contributions exist even if the chargino is not a pure Higgsino.
We then obtain
\beq
\delta C_7 \simeq \sgn (\mu ) 
               \frac{1}{6}\frac{m_t}{m_{\tilde H^\pm}} T_{11}T_{12}\tan\beta
              J\left( \frac{m^2_{\tilde t_1}}{m^2_{\tilde H^\pm}}\right)\,.
\eeq
Here it is crucial that there be substantial $\tilde t_R$ and $\tilde t_L$ 
contributions to $\tilde t_1$.  As argued above, this is generic in minimal
supergravity theories.  This expansion demonstrates that 
large $\tan\beta$ solutions ($\tan\beta \gsim 40$)
can yield greater than ${\cal O}(1)$ contributions to
$R_7$ for mass scales of $1\tev$.  Even rather low values of $\tan\beta$
exhibit enhancements with a light Higgsino and light mixed stop.  This
is demonstrated for $\tan\beta =2$ in Fig.~\ref{fig11} (solid line).
Furthermore, large contributions are
possible in both the negative and positive directions of $R_7$ depending on the
sign of $\mu$. For example, with $m_{\tilde t_1}=250\gev$,
$|T_{11}T_{12}|=1/10$,
$m_{\tilde \chi^\pm_1}=250\gev$ and $\tan\beta =50$ we find
that $|R_7(M_W)| \gsim 3$.
Again, we are in a region of 
parameter space which is not tuned
just to give this large effect in $B\to X_s\gamma$, rather we are in a region
which is highly motivated by $SO(10)$ grand unified theories.  
The finite corrections to the $b$ quark mass~\cite{hall,pierce} constitute
approximately a $20\%$ correction to the $b$ Yukawa coupling when 
$\tan\beta \gsim 40$.  This $b$ Yukawa coupling
is implicitly present in $\lambda^R_i$. Depending on the sign of $\mu$ this
correction can be positive or negative.  We don't include these finite
$b$ mass corrections
in our analysis since it requires a detailed knowledge of the sparticle
spectrum which we are not specifying.

The chirality structure of the $O_i$ operators allow 
a large $\tan\beta$ enhancement only for the $C_{7,8}(\mw )$ coefficients.
In $O_7$ the $b_R$ quark must undergo an helicity flip as long as we neglect
$m_s$ dependent effects.  Therefore, all contributions to $C_7$ 
in the standard model
and MSSM must be proportional to the $b$-quark mass.  However,
some diagrams with $\tilde t_k/\tilde \chi^\pm_j$ loops allow proportionality
to the bottom Yukawa alone, which yields a $m_b/\cos\beta$ enhancement for
large $\tan\beta$~\cite{okada93:119}.  
The mixings between the charged Higgs weak eigenstate
and goldstone forbid $m_b/\cos\beta$ enhancements in the physical
charged Higgs graphs.  Furthermore, the helicity structure of the
four-fermion operators $O_{9,10}$ forbid large $\tan\beta$ enhancements.

We conclude our analysis by examining the charged Higgs contributions 
to $B\to X_s\gamma$ alone.  It is 
well-known~\cite{hewett:94,rizzo:88} 
that a $H^\pm$ boson can contribute significantly
to $C_{7,8}$, but has a smaller effect on the coefficients $C_{9,10}$;
this is also illustrated above in Figs.~8,9,11.  
The form of the coefficients of 
the magnetic dipole operators in this case are presented in the appendix.
From these equations, 
we see that not only do large enhancements occur for small 
values of $\tan\beta$, but more importantly, the coefficients are always 
larger than those of the SM, independent of the value of $\tan\beta$.  This 
leads to the familiar bound $m_{H^\pm}>260\gev$ obtained from the measurement 
of $B(\bsg)$ by CLEO.  
However, this constraint does not make use of the recent NLO calculation of 
the matrix elements for this decay which are discussed in previous sections.
We remind the reader that a full NLO calculation would also require the
higher order matching conditions for the SM and $H^\pm$ 
contributions as well as
the NLO anomalous dimensions for $C^{eff}_7(\mu)$.  
Nevertheless, we recall that
preliminary results on the NLO corrections to $C^{eff}_7(\mu)$ 
indicate they are
small\cite{misiak:96}, and a good approximation is obtained by employing
the leading order expression for $C^{eff}_7(\mu)$ with the NLO matrix elements.
Since this drastically reduces the $\mu$ dependence of the branching fraction,
we would expect the $H^\pm$ constraints to improve.  Indeed, we find that
the CLEO bound excludes the region to the left and beneath the curves in
Fig.~\ref{fig12}.  For $m_t^{phys}=169\gev$ we see that $m_{H^\pm}>300\gev$.
This is calculated by using the same procedure that produced the previous
charged Higgs mass bound by CLEO,  \ie, all the input parameters (\eg, 
$\alpha_s$,
$\mu$, $m_c/m_b$, and $B(B\to X\ell\nu)$) are varied over their allowed ranges 
in order to ascertain the most conservative limit.  This bound holds in
the general two-Higgs-doublet-model II, and in supersymmetry if the
superpartners are all significantly massive.  

\jfig{fig12}{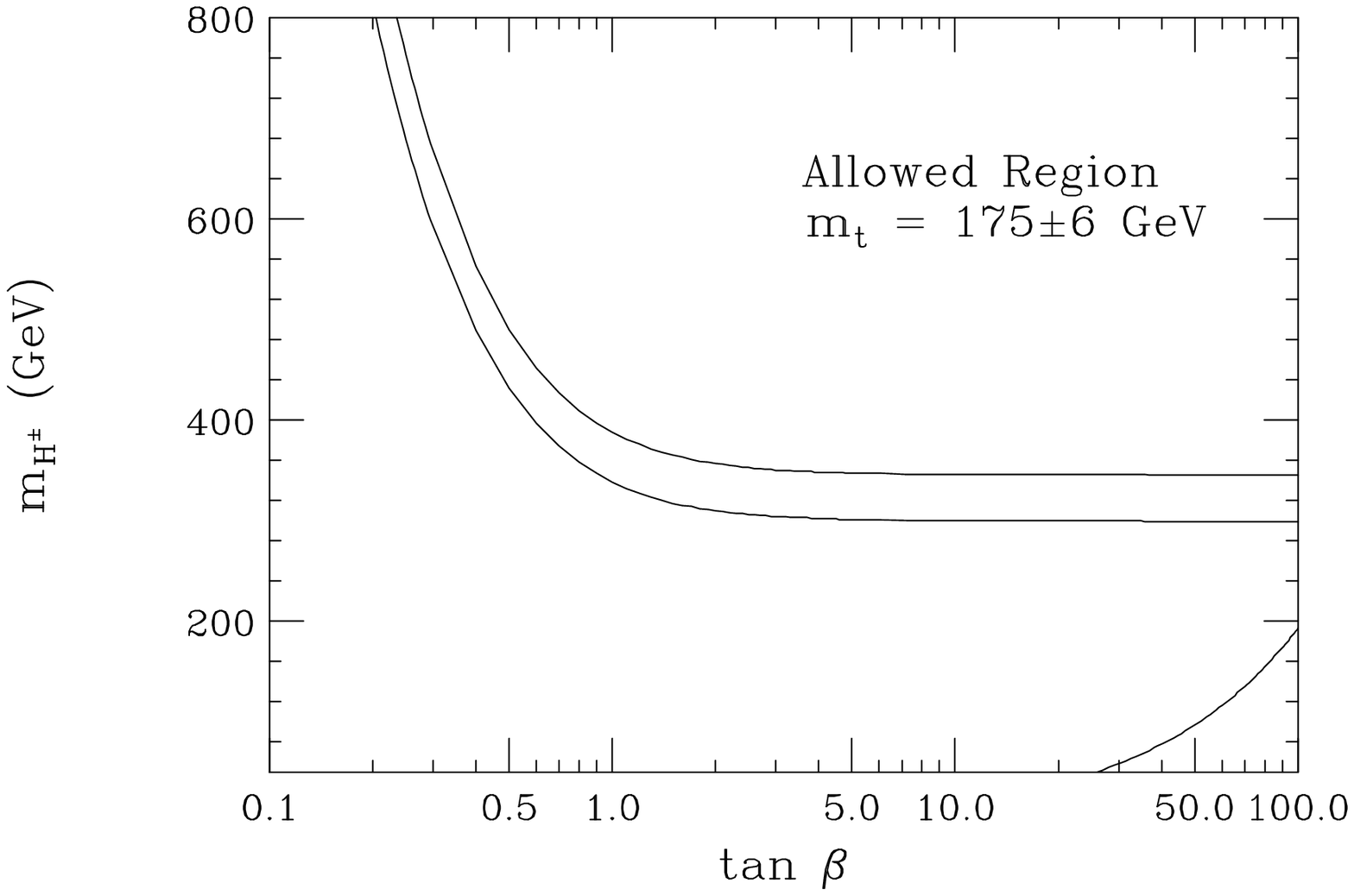}{Constraints in the charged Higgs mass -- $\tan\beta$
plane from the CLEO bound on $B(\bsg)$.  The excluded region is that to the
left and below the curves. The top line is for $m_t^{phys}=181\gev$
and the bottom line is for $m_t^{phys}=169\gev$.
We also display the restriction 
$\tan\beta /m_{H^\pm}>0.52\gev^{-1}$ which arises from measurements
of $B\to X\tau\nu$ as discussed in~\protect\cite{grossman}.}

\section{Conclusion}

In this paper we have studied the effects of supersymmetry
to the FCNC observables concerning $b\to s$ transitions, 
and we have seen that deviations from
the standard model could be detected with supersymmetric masses 
even at the TeV scale.  This is especially true if $\tan\beta$ is very high.
The large $\tan\beta$ enhancements in the $b\to s$ processes are unique.  
Deviations in $B-\bar B$ mixing are not as pronounced
since the $m_b/\cos\beta$ enhancements, which are possible in this case, 
must be compared
with $m_t$, whereas in the magnetic dipole $b\to s$ transitions the
$m_b/\cos\beta$ enhancement is to be compared with $m_b$.
Furthermore, the uncertainties in the decay constant and bag factor make
$B-\bar B$ mixing observables slightly less appealing when trying to
probe deviations from the standard model.  

Therefore, it is possible that the first distinct signs
of supersymmetry could come from deviations in $b\to s$ transitions.
One would like direct confirmation of a possible deviation at the
$B$-factory, and collider programs such as an upgraded luminosity
Tevatron~\cite{upgrade}, 
and the LHC could provide it.  At the Tevatron, the cleanest
signal for supersymmetry is the $3l$ signal coming from leptonic decays
of $p\bar p\to \chi^\pm_1\chi^0_2$.  If the light charginos and neutralinos
are mostly gaugino-like then the branching fraction into $3l$ can be quite
high.  This is true when the sleptons are lighter than the squarks and
near $M_W$, as the slepton mediated decays of the gauginos enhance the final
state leptons branching fraction.  If the charginos and neutralinos contained
a substantial Higgsino fraction then the slepton mediated exchanges are 
suppressed by lepton Yukawas to a negligible level, and all the decays must
proceed through $W$ and $Z$ bosons, and recall that the branching fraction of
$WZ\to l^+l^-l'\lsim 2\%$ (not counting $\tau$'s.) 
On the other hand, the $b\to s$ transitions are largest when there is a
substantial Higgsino fraction in the lightest chargino state.  
The good news is that the
trilepton signal and $b\to s$ decays are somewhat complementary in
supersymmetric parameter space.  The bad news is that confirmation between
the two experiments might be difficult.  Confirmation could be possible
later at the LHC  through total missing energy signatures, or at the NLC
through direct mass and mixing angle measurements which would allow 
SUSY predicted $b\to s$ rates to be compared with the data.

Much of the signal phenomenology at high energy colliders such as the Tevatron
and LHC rely on missing energy signatures.  However, these are only possible
with a stable LSP, which is the result of a postulated R-parity among the
fields.  R-parity conservation certainly need not hold in the correct theory
of nature, and even tiny R-parity violating couplings destroy the missing
energy signatures.  More complicated analyses then must be performed with
greatly reduced sensitivity to sparticle production. The Tevatron upgrade
then would have difficulty exceeding equivalent
LEPII bounds on sparticle masses~\cite{rtevatron}.  The LHC
would have a significant search capability beyond LEPII, although much
reduced compared to search capabilities with R-parity conservation if
the LSP decays hadronically~\cite{rlhc}.
It appears to be difficult to extract SUSY signals at the LHC for superpartners
above $1\tev$ in this case.
Searches for virtual
sparticles, such as those we discuss in this paper, do not suffer in the
presence of R-parity violation, and we have already noted that SUSY 
contributions are resolvable with masses above $1\tev$.  
In fact, the signal may be enhanced over
the rate for gauge interactions alone if
the R-parity violating
couplings are sufficiently large.   It is therefore possible that the
$B$-factory could be sensitive to some parts in supersymmetric parameter
space not accessible at the LHC.  

\bigskip
\noindent
{\it Acknowledgements.} We would like to thank C.~Greub, Y.~Grossman,
A.~Kagan, T.~Rizzo, and M.~Worah for helpful discussions.

\section*{Appendix}

In this appendix all the standard model and supersymmetric
contributions to the matching conditions $C_{7-10}(\mw )$ are listed
in the limit
that all squarks are degenerate except the stop squarks.  It is also assumed
that no sources of flavor violation are allowed other than the CKM angles.

The dipole moment operators $C_{7,8}(\mw )$ are already provided in the
literature in several places for this limit.  For completeness
we write them down here using the formulas of~\cite{barbieri93:86} which
are normalized according to our definition above:
\bea
C_{7,8}^W(\mw ) & = & \frac{3}{2}x_W\fyg{1}(x_W), \\
C_{7,8}^{H^\pm}(\mw ) & = & \frac{x_H}{2}\left[
   \frac{1}{\tan^2\beta}\fyg{1}(x_H)+\fyg{2}(x_H)\right], \\
C_{7,8}^{\chi^\pm}(\mw ) & = & 
      \sum_i \frac{\mw^2}{\mxc{i}^2} V_{i1}^2 \fyg{1}(y_{\tilde q i})-
      \sum_{i,k} \frac{\mw^2}{\mxc{i}^2} \Lambda_{ik}^2 \fyg{1}(y_{ki}) \\
   & & -\sum_i\frac{U_{i2}}{\sqrt{2}\cos\beta}\frac{\mw}{\mxc{i}}
        V_{i1}\fyg{3}(y_{\tilde q i})
       +\sum_{i,k}\frac{U_{i2}}{\sqrt{2}\cos\beta}\frac{\mw}{\mxc{i}}
        \Lambda_{ik}T_{k1}
      \fyg{3}(y_{ki})
\eea
where
\bea
x_W & = &\frac{m_t^2}{\mw^2},~~x_H=\frac{m_t^2}{m_{H^\pm}^2},~~
y_{\tilde q i}=\frac{m^2_{\tilde q}}{\mxc{i}^2},~~
y_{k i}=\frac{m^2_{\tilde t_k}}{\mxc{i}^2}, \\
\Lambda_{ik} & = & V_{i1}T_{k1}-V_{i2}T_{k2}\frac{m_t}{\sqrt{2}\mw\sin\beta}
\eea
and
\bea
\fy{1}(x) & = & \frac{7-5x-8x^2}{36(x-1)^3}+\frac{x(3x-2)}{6(x-1)^4}\log x, \\
\fy{2}(x) & = & \frac{3-5x}{6(x-1)^2}+\frac{3x-2}{3(x-1)^3}\log x, \\
\fy{3}(x) & = & (1-x)\fy{1}(x)-\frac{x}{2}\fy{2}(x)-\frac{23}{36}, \\
\fg{1}(x) & = & \frac{2+5x-x^2}{12(x-1)^3}-\frac{x}{2(x-1)^4}\log x, \\
\fg{2}(x) & = & \frac{3-x}{2(x-1)^2}-\frac{1}{(x-1)^3}\log x, \\
\fg{3}(x) & = & (1-x)\fg{1}(x)-\frac{x}{2}\fg{2}(x)-\frac{1}{3}.
\eea

It is convenient to write $C_{9,10}(\mw )$ as
\bea
C_9(\mw ) & = & \frac{Y-4Z\sin^2\theta_W}{\sin^2\theta_W}~~\hbox{\rm and} \\
C_{10}(\mw ) & = & \frac{-Y}{\sin^2\theta_W}.
\eea
The values for $Y$ and $Z$ are contained in~\cite{cho96:360} for arbitrary
flavor structure and masses.
For all squarks degenerate except the stop squarks and with only CKM flavor
violation then
\bea
Y & = & Y_t+Y^Z_{H^\pm}+Y^\gamma_{H^+}+Y^{Z}_{\chi^\pm}+Y^\gamma_{\chi^\pm}
       +Y^{box}_{\chi^\pm}, \\
Z & = & Z_t+Z^Z_{H^\pm}+Z^\gamma_{H^+}+Z^{Z}_{\chi^\pm}+Z^\gamma_{\chi^\pm}
       +Z^{box}_{\chi^\pm}.
\eea
The functional form of each of these contributions is
\bea
Y_t & = & \frac{4x_W-x_W^2}{8(1-x_W)}+\frac{3x_W^2}{8(1-x_W)^2}\log x_W, \\
Z_t & = & \frac{108x_W-259x_W^2+163x_W^3-18x_W^4}{144(1-x_W)^3} \\
    & &   +\frac{-8+50x_W-63x_W^2-6x_W^3+24x_W^4}{72(1-x_W)^4}\log x_W, \\
Y^Z_{H^+}=Z^Z_{H^+} & = & -\frac{1}{8}\cot^2\beta x_W f_5(x_H), \\
Y^\gamma_{H^+} & = & 0, \\
Z^\gamma_{H^+} & = & -\frac{1}{72}\cot^2\beta f_6 (x_H), \\
Y^Z_{\chi^\pm} = Z^Z_{\chi^\pm} & = &
    \sum_i -V_{i1}^2 L_1(\mxc{i}^2,\msq^2,\msq^2)
   +\sum_{i,k,l} \Lambda_{ki}\Lambda_{kl} T_{k1}T_{l1}
      L_1(\mxc{i}^2,\mst{k}^2,\mst{l}^2) \\
   & & + \sum_{i,j} -V_{i1}V_{j1}L_2(\msq^2,\mxc{i}^2,\mxc{j}^2)
 + \sum_{i,j,k} 
     \Lambda_{ki}\Lambda_{kj}L_2(\mst{k}^2,\mxc{i}^2,\mxc{j}^2), \\
Y^\gamma_{\chi^\pm} & = & 0, \\
Z^\gamma_{\chi^\pm} & = & \sum_i -V_{i1}^2
     L_3(\mxc{i}^2,\msq^2)+\sum_{i,k}\Lambda_{ki}^2
             L_3(\mxc{i}^2,\mst{k}^2), \\
Y^{box}_{\chi^\pm} & = &  \sum_{i,j} -V_{i1}V_{j1}
  L_4(\mxc{i}^2,\mxc{j}^2,\msq^2,m_{\tilde \nu}^2)+
  \sum_{i,j,k} \Lambda_{ki}\Lambda_{kj} 
  L_4(\mxc{i}^2,\mxc{j}^2,\mst{k}^2,m_{\tilde \nu}^2), \\
Z^{box}_{\chi^\pm} & = & 0.
\eea
In the above equations $\msq$ is the common squark mass.  The functions
are defined as
\bea
L_1(\mxc{i}^2,m_1^2,m_2^2) & = & \frac{1}{2}c_2(\mxc{i}^2,m_1^2,m_2^2) \\
L_2(m^2,\mxc{i}^2,\mxc{j}^2) & = & -\frac{1}{2}c_2(m^2,\mxc{i}^2,\mxc{j}^2)
  V_{i1}V_{j1} \\
    &  & +\frac{1}{4}\mxc{i}\mxc{j}c_0(m^2,\mxc{i}^2,\mxc{j}^2)
  U_{i1}U_{j1}, \\
L_3(\mxc{i}^2,m^2) & = & \frac{\mw^2}{36m^2}f_7(\mxc{i}^2/m^2), \\
L_4(\mxc{i}^2,\mxc{j}^2,\msq^2,m_{\tilde \nu}^2) & = & \mw^2
  d_2(\mxc{i}^2,\mxc{j}^2,\msq^2,m_{\tilde \nu}^2)V_{i1}V_{j1}.
\eea
These $L$-functions are expressed in terms of functions contained 
in~\cite{cho96:360} and are explicitly given by
\bea
f_5(x) & = & \frac{x}{1-x}+\frac{x}{(1-x)^2}\log x, \\
f_6(x) & = & \frac{38x-79x^2+47x^3}{6(1-x)^3}+\frac{4x-6x^2+3x^4}{(1-x)^4}
              \log x, \\
f_7(x) & = & \frac{52-101x+43x^2}{6(1-x)^3}+\frac{6-9x+2x^3}{(1-x)^4}
             \log x, \\
c_0(m_1^2,m_2^2,m_3^2) & = &
   -\Bigg[ \frac{m_1^2\log\frac{m_1^2}{\mu^2}}{(m_1^2-m^2_2)(m_1^2-m_3^2)}
       + (m_1 \leftrightarrow m_2 ) + (m_1\leftrightarrow m_3)\Bigg], \\
c_2(m_1^2,m_2^2,m_3^2) & = &
 \frac{3}{8}-\frac{1}{4}
 \Bigg[ \frac{m_1^4\log\frac{m_1^2}{\mu^2}}{(m_1^2-m^2_2)(m_1^2-m_3^2)}
       + (m_1 \leftrightarrow m_2 ) + (m_1\leftrightarrow m_3)\Bigg], \\
d_2(m^2_1,m^2_2,m^2_3,m^2_4) & = &
 -\frac{1}{4}\Bigg[ 
\frac{m^4_1\log
   \frac{m^2_1}{\mu^2}}{(m^2_1-m^2_2)(m^2_1-m^2_3)(m_1^2-m_4^2)}  \\
  & & ~~~~+ (m_1 \leftrightarrow m_2 ) 
     + (m_1\leftrightarrow m_3)+ (m_1 \leftrightarrow m_4 )\Bigg]  .
\eea



\begin{thebibliography}{20}

\bibitem{buraswarsaw} A.~Buras, Plenary talk given at the 
{\it 28th International Conference on High Energy Physics}, Warsaw, Poland,
July 1996.

\bibitem{buras94:374} A.~Buras, M. Misiak, M.~M\" unz,
S.~Pokorski, \NPB{424}{94}{374}.

\bibitem{greub:96} C. Greub, T. Hurth, and D. Wyler, \PLB{380}{96}{385};
\PRD{54}{96}{3350}; C. Greub and T. Hurth, talk presented at DPF96,
Minneapolis, MN, August 1996, hep-ph/9608449.

\bibitem{buras:95} A.J. Buras and M. M\" unz, \PRD{52}{95}{186}.

\bibitem{misiak:96} K.G. Chetyrkin, M. Misiak, and M. M\" unz, in preparation;
M. Misiak, talk presented at ICHEP96, Warsaw, Poland, July 1996.

\bibitem{cleo:94} CLEO Collaboration, M.S. Alam \etal, \PRL{74}{95}{2885}.

\bibitem{bsll} CLEO Collaboration, R.~Balest \etal, in {\it Proceedings
of the 27th International Conference on High Energy Physics}, Glasgow,
Scotland, 1994, edited by P.J.~Bussey and I.G.~Knowles (IOP, London,
1995); CDF Collaboration, C.~Anway-Wiese, in {\it The Albuquerque Meeting},
Proceedings of the 8th Meeting of the Division of Particles and Fields of the
American Physical Society, Albuquerque, New Mexico, 1994, edited by S.~Seidel
(World Scientific, Singapore, 1995).

\bibitem{ali} A. Ali, T. Mannel, and T. Morozumi, \PLB{273}{91}{505}.

\bibitem{ali2}
A. Ali, G.F. Giudice, and T. Mannel, \ZPC{67}{95}{417}.

\bibitem{hewett} J.L.~Hewett, \PRD{53}{96}{4964}.

\bibitem{grinstein:89} See, for example, B. Grinstein, M.J. Savage, and M.
Wise, \NPB{319}{89}{271}; A. Ali, in {\it 20th International Nathiagali
Summer College on Physics and Contemporary Needs}, Bhurban, Pakistan, 1995,
hep-ph/9606324.

\bibitem{inami} T. Inami and C.S. Lim, Prog. Theor. Phys. {\bf 65}, 297
(1981).

\bibitem{pdg} R.M. Barnett, \etal, (Particle Data Group), \PRD{54}{96}{1}.

\bibitem{schmelling} M.~Schmelling, Plenary talk given at the 
{\it 28th International Conference on High Energy Physics}, Warsaw, Poland,
July 1996.

\bibitem{cdfd0} P. Tipton, talk presented at {\it 28th International Conference
on High Energy Physics}, Warsaw, Poland, July 1996.

\bibitem{richman} J. Richman, Plenary talk given at the 
{\it 28th International Conference on High Energy Physics}, Warsaw, Poland,
July 1996.

\bibitem{desh:89} N.G. Deshpande, J. Trampetic, and K. Panrose, 
\PRD{39}{89}{1461}; C.S. Lim, T. Morozumi, and A.I. Sanda, \PLB{218}{89}{343};
Z.~Ligeti and M.~Wise, \PRD{53}{96}{4937}.

\bibitem{greub:91} A. Ali and C. Greub, \ZPC{49}{91}{431}; \PLB{259}{91}{182};
\PLB{361}{95}{146}; N. Pott, \PRD{54}{96}{938}.

\bibitem{yao} K. Adel and Y.-P. Yao, \PRD{49}{94}{4945}.

\bibitem{hqet} J.~Chay, H.~Georgi, and B.~Grinstein, \PLB{247}{90}{399};
I.~Bigi, N.~Uraltsev, and A.~Vainshtein, \PLB{243}{92}{430};
B.~Blok and M.~Shifman, \NPB{399}{93}{441};
I.~Bigi~\etal , \PRL{71}{93}{496}.

\bibitem{nir93:337} Y.~Nir, N.~Seiberg, \PLB{309}{93}{337}.

\bibitem{dine} M.~Dine, A. Nelson, \PRD{48}{93}{1277}; M.~Dine,
A.~Nelson, Y.~Shirman, \PRD{51}{95}{1362}.

\bibitem{dimo1} S.~Dimopoulos, S.~Thomas, J.D.~Wells, hep-ph/9609434.

\bibitem{dimo2} S.~Dimopoulos, H.~Georgi, \NPB{193}{81}{150}.

\bibitem{rareb}
S. Bertolini, F.~Borzumati, A.~Masiero, G.~Ridolfi, \NPB{353}{91}{591}.
F. Borzumati, \ZPC{63}{94}{291};
F. Gabbiani, E. Gabrielli, A. Masiero, L. Silvestrini, hep-ph/9604387;
V. Barger, M.S. Berger, P. Ohmann, R.J.N. Phillips, \PRD{51}{95}{2438};
D. Choudhury, F. Eberlein, A. Konig, J. Louis, S. Pokorski, \PLB{342}{95}{180};
J. Lopez, D. Nanopoulos, X. Wang, A.~Zichichi, \PRD{51}{95}{147}.

\bibitem{cho96:360}P. Cho, M. Misiak, D. Wyler, hep-ph/9601360.

\bibitem{kkrw}For description of procedure we follow, see 
G.L.~Kane, C.~Kolda, L.~Roszkowski, J.~Wells, \PRD{49}{94}{6173}.

\bibitem{anlauf94:245} We are neglecting potentially important
contributions coming from disparate SUSY and electroweak scales, for a
discussion of this see H. Anlauf, \NPB{430}{94}{245}.

\bibitem{nelson} L.~Randall, R.~Sundrum, \PLB{312}{93}{148}; B.~Grinstein,
Y.~Nir, and J.M.~Soares, \PRD{48}{93}{3960};
A.~Nelson, M.~Strassler, hep-ph/9607362.

\bibitem{Rbpapers}M.~Boulware, D.~Finnell, \PRD{44}{91}{2054};
J.D.~Wells, C.~Kolda, G.L.~Kane, \PLB{338}{94}{219}.

\bibitem{transition} M.~Carena, M.~Quiros, C.~Wagner, \PLB{380}{96}{81}.

\bibitem{hk} H. Haber, G.L. Kane, Phys. Rep. 142B (1984) 212.

\bibitem{brignole96:293}A. Brignole, F. Feruglio, F. Zwirner,
hep-ph/9601293.

\bibitem{garisto93:372}R. Garisto, J. Ng, \PLB{315}{93}{372}.

\bibitem{hall}L.~Hall, R.~Rattazzi, U.~Sarid, \PRD{50}{94}{7048}.

\bibitem{pierce}D.~Pierce, J.~Bagger,K.~Matchev, R.~Zhang,
hep-ph/9606211.

\bibitem{okada93:119} Y. Okada, \PLB{315}{93}{119}.

\bibitem{hewett:94} J.L. Hewett, \PRL{70}{93}{1045}, V. Barger, M. Berger,
and R.J.N. Phillips, \PRL{70}{93}{1368}; V. Barger, J. Hewett, R. Phillips,
\PRD{41}{90}{3421}; N.~Deshpande, K.~Panose, J.~Trampetic,
\PLB{308}{93}{322}.

\bibitem{rizzo:88} T.G. Rizzo, \PRD{38}{88}{820}; X.G. He \etal,
\PRD{38}{88}{814}; W.S. Hou and R.S. Willey, \PLB{202}{88}{591};
C.Q. Geng and J.N. Ng, \PRD{38}{88}{2858}; B. Grinstein, R. Springer, and
M. Wise, \NPB{339}{90}{269}.

\bibitem{grossman} P.~Krawczyk and S.~Pokorski, \PRL{60}{88}{182};
Y.~Grossman, H.~Haber, and Y.~Nir, \PLB{357}{95}{630}.

\bibitem{upgrade} H.~Baer, C.-h. Chen, C.~Kao, X.~Tata, \PRD{52}{95}{1565};
S.~Mrenna, G.L.~Kane, G.~Kribs, J.D.~Wells, \PRD{53}{96}{1168}.

\bibitem{rtevatron} D.P.~Roy, \PLB{283}{92}{270}; H.~Baer, C.~Kao,
X.~Tata, \PRD{51}{95}{2180}.

\bibitem{rlhc} H.~Baer, C.-h.~Chen, X.~Tata, hep-ph/9608221.

\bibitem{barbieri93:86}R. Barbieri, G. Giudice, \PLB{309}{93}{86}.

\end{thebibliography}
\end{document}